\documentclass[]{aastex7}

\usepackage{soul}
\usepackage{graphicx}
\usepackage{hyperref}

\submitjournal{ApJ}

\begin{document}

\title{Magnetic Field Variability as a Consistent Predictor of Solar Flares}






\author[0000-0003-3740-9240]{K. L. Kniezewski}
\affiliation{Air Force Institute of Technology \\
2950 Hobson Way \\
Wright-Patterson AFB, OH 45433, USA}
\email{kara.kniezewski@us.af.mil}

\author[0000-0002-8767-7182]{E. I. Mason}
\affiliation{Predictive Science Inc. \\
9990 Mesa Rim Rd, Suite 170 \\
San Diego, CA 92121, USA}
\email{emason@predsci.com}

\author[0000-0002-3495-6372]{D. J. Emmons}
\affiliation{Air Force Institute of Technology \\
2950 Hobson Way \\
Wright-Patterson AFB, OH 45433, USA}
\email{daniel.emmons.1@us.af.mil}

\author[0000-0003-3322-9955]{K. E. Fitch}
\affiliation{Air Force Institute of Technology \\
2950 Hobson Way \\
Wright-Patterson AFB, OH 45433, USA}
\email{kyle.fitch@us.af.mil}

\author[0000-0002-9359-6776]{S. H. Garland}
\affiliation{Air Force Institute of Technology \\
2950 Hobson Way \\
Wright-Patterson AFB, OH 45433, USA}
\email{seth.garland.1@us.af.mil}

\begin{abstract}
Solar flares are intense bursts of electromagnetic radiation, which occur due to a rapid destabilization and reconnection of the magnetic field. While pre-flare signatures and trends have been investigated from magnetic observations prior to flares for decades, analysis which characterizes the variability of the magnetic field in the hours prior to flare onset has not been included in the literature. Here, the 3D magnetic field is modeled using a Non-Linear Force Free Field extrapolation for 6 hours before and 1 hour after 18 on-disk solar flares and flare quiet windows for each active region. Parameters are calculated directly from the magnetic field from two field isolation methods: the ``Active Region Field", which isolates field lines where the photospheric field magnitude is $\geq$ 200 Gauss, and the ``High Current Region", which isolates field lines in the 3D field where the current, non-potential field, twist, and shear exceed pre-defined thresholds. For this small pool of clean events, there is a significant increase in variation starting 2-4 hours before flare onset for the current, twist, shear, and free energy, and the variation continues to increase through the flare start time. The current, twist, shear, and free energy are also significantly stronger through the lower corona, and their separation from flare quiet height curves scales with flare strength. Methods are proposed to combine variation of the magnetic fields with variation of other data products prior to flare onset, suggesting a new potential flare prediction capability.
\end{abstract}


\section{Introduction} \label{sec:intro}
Solar flares are an impulsive release of electromagnetic radiation over a wide range of wavelengths from x-rays ($10^{-4}$ nm) through the radio spectrum ($10^{11}$ nm) \cite[e.g.,][]{Toriumi2019, Tobiska2004}. Solar flares are often sources of several radiation events; a few events which have notable impact on Earth are solar energetic particles \cite[SEPs; e.g.,][]{Reames1990}, various spectra of radiation \cite[specifically, ultraviolet (UV) emission;][]{Lean1997}, and coronal mass ejections \cite[CMEs; e.g.,][]{Andrews2003}. X-ray and extreme ultraviolet (XUV/EUV) flux, which originate at high temperatures from the chromosphere through the corona and increase with solar activity, can cause large disturbances to satellite radio communications due to an enhanced ionosphere \citep{Klobuchar1985} and atmospheric drag on satellites due to increased atmospheric density, dramatically shortening their operational lifetime \citep{DeLafontaine1982, Pulkkinen2007}. SEPs arrive at Earth with near-relativistic energies and are capable of piercing the skin of spacecraft and instruments. Coronal mass ejections (CMEs), when directed at Earth, a CME can cause geomagnetic storms within 1–4 days \citep{Schwenn2006}. Due to the severe nature of these solar events on technology, studies focused on creating forecasting and early warning tools have become crucial as technological dependence continues to grow.

The solar flare standard model \citep{Carmichael1964, Sturrock1968, Hirayama1974, Kopp1976} provides a framework for magnetic field evolution, which describes their onset and energy release properties. It is generally accepted that the primary mechanism for energy release in solar flares is magnetic reconnection in the Sun's corona. Thoroughly described in \cite{Holman2016}, photospheric motions such as differential rotation and/or granular flow cause magnetic loops, which connect the two polarity regions, to elongate due to shearing. This process causes these highly-sheared loops to become unstable, collapse inward, and reconnect with one another. As long as reconnection continues, upward ejected field builds a magnetic flux rope above the reconnection site. If this magnetic flux rope becomes unstable, then it can eject as a coronal mass ejection. While strong flares are typically associated with CMEs \citep{Yashiro2006}, solar flares are not always accompanied by a CME, and CMEs can come from several eruptive phenomena, such as prominence eruptions \citep{Gosling1993}. Various theories and models have been proposed for field destabilization to initiate an eruption \cite[see][for a review]{Green2018}; major areas of active research involve the ideal instabilities \cite[torus and kink instabilities;][]{Kliem2004, Torok2004}, which investigate the stability of a magnetic flux rope and the strength of the overlying field, and the magnetic breakout model \citep{Antiochos1999a}, which concerns the overall magnetic complexity of the erupting field. 

The goal of solar flare forecasting is to answer the following question: when and where will a solar flare occur on the solar disk, and how strong will it be? Answering this question, however, has been quite challenging since it is currently not attainable to directly observe the optically-thin magnetic field in the low-density and high-temperature environment of the corona. We can measure the photospheric magnetic field, which is currently used as the primary data product to search for trends or consistent signatures prior to solar flare \cite[e.g.,][and many others]{Leka2003, Schrijver2007, Bobra2015, Kazachenko2022}. While studies utilizing the photospheric field draw intriguing statistical relations to different magnetic field properties and solar flare likelihood, these methods have not been able to uncover consistent behavior from the photospheric magnetic field observations. Since solar flares are accepted to occur in the low corona, there has recently been increasing work in modeling the magnetic field in the solar corona by using the photospheric field as a boundary condition in order to discover the underlying mechanisms of flares \cite[e.g.,][]{Gupta2021, Yurchyshyn2022, Garland2024}. Non-linear force-free field (NLFFF) modeling methods have faster computation times compared to more sophisticated models, and, while containing a list of assumptions, are a common model for flare prediction studies. 

Ultimately, the majority of studies utilizing pre-flare magnetic field observations encounter challenges when searching for consistent trends or signatures in the temporal variation of the magnetic fields. In a recent study, \cite{Kniezewski2024arxiv} investigated the off-limb extreme ultraviolet (EUV) emission from quiescent coronal loops prior to over 50 solar flares, finding that the pre-flare emission is significantly more variable than flare quiet conditions. There is no consistent trend, however, in how the emission varies prior to, nor is there correlation between EUV channels. Consequently, analysis of the EUV emission is focused on characterizing the variation of the emission in terms of time before flares in comparison to flare quiet conditions. This form of analysis yields important implications for flare strength, which may be useful for flare prediction methods. Here, we utilize similar forms of analysis on the magnetic fields to search for consistent patterns of magnetic variation prior to flare onset.

In this paper, we extrapolate the coronal magnetic field using a NLFFF model for 6 hours before and 1 hour after 18 flares to focus on characterizing magnetic changes in the corona prior to flare onset, rather than searching for trends or signatures in the time series themselves. We develop 2 techniques to isolate specific portions of the magnetic field with automated selection methods and low computation time. The magnetic field prior to solar flares has been investigated for decades; we propose that the analysis methods adapted from \cite{Kniezewski2024arxiv} for the magnetic fields can be utilized to predict significant flares with hours of warning by searching for when are variations in the magnetic field most significant, and how strong their variation is compared to flare quiet conditions. 

Section \ref{sec:data_methods} contains a detailed description of our data selection, coronal magnetic field modeling techniques, calculations, and magnetic field isolation methods. Section \ref{sec:results} discusses the results and analysis of the pre-flare coronal loop models. Implications and conclusions are discussed in Section \ref{conclusions}.

\section{Data Selection and Methods}\label{sec:data_methods}
Since it is not currently feasible to directly measure the coronal magnetic field due to observational limitations, there is significant benefit to modeling the coronal field to gain insight into solar flare onset and energy release. Specifically, the purpose of utilizing coronal field modeling in this work is to provide information about the three-dimensional magnetic field volume which is far more extended into the solar atmosphere prior to solar eruptions.

\subsection{Data Selection}\label{sec:data}
Flares were selected from a list of events analyzed in \cite{Mason2022}, which is freely available at \cite{rain_dataset}. This flare list contains 241 flares between January 2011 and December 2019, which has daily coverage from the Solar Dynamics Observatory (SDO) and covers most of Solar Cycle 24. All X-class flares from Solar Cycle 24 are included, as well as a distribution of C- and M-class flares which distribute throughout Solar Cycle 24 and in magnitude. Flares were chosen based on limiting conditions to ensure the data volumes from the active region of study had minimal saturation from other events that occurred previously. Specifically, flares were removed if an active region had a C5.0 flare or higher occur within six hours of an event of interest within the same active region. Flares were also only selected if they were within $\pm$ 45 degrees off disk-center to minimize projection effects. Time windows when the selected flaring active regions were not flaring (no flares $\geq$ C5.0 for a 7 hour period) were also recorded to compare pre-flare signatures to quiet-time signatures. The selected 18 flares, which comprise 6 C-Class, 6 M-Class, and 6 X-class, and their corresponding active region flare quiet time windows are listed in Table \ref{tab:events}.

\renewcommand{\arraystretch}{1.5}
\begin{table}[h!]
    \centering
    \caption{Flares and corresponding AR flare-quiet window used in this study.}
    \begin{tabular}{c c c c c c}
    \hline
    \textbf{GOES Class} & \textbf{Start Time (UT)} & \textbf{NOAA AR} & \textbf{CME} & \textbf{Hale Class} & \textbf{Flare Quiet Start (UT)} \\
    \hline
    X3.0 & 2011-09-06 22:12 & 11283 & Yes & $\beta\gamma$ & 2011-09-05 12:00 \\
    X2.7 & 2014-12-20 00:11 & 12242 & Yes & $\beta\gamma\delta$ & 2014-12-18 14:00 \\
    X2.4 & 2014-09-10 17:21 & 12158 & Yes & $\beta\gamma\delta$ & 2014-09-09 15:00 \\
    X1.6 & 2013-11-08 04:20 & 11890 & Yes & $\beta\gamma\delta$ & 2013-11-08 15:00 \\
    X1.5 & 2014-03-29 17:35 & 12017 & Yes & $\beta\delta$ & 2014-03-28 00:30 \\
    X1.1 & 2015-06-25 08:02 & 12371 & Yes & $\beta\gamma$ & 2015-06-23 17:00 \\
    M9.1 & 2013-11-01 19:46 & 11884 & Yes & $\beta\gamma\delta$ & 2013-10-31 09:00 \\
    M8.1 & 2012-07-02 10:43 & 11515 & Yes & $\beta\gamma$ & 2012-07-01 19:00 \\
    M4.6 & 2013-05-17 08:43 & 11748 & Yes & $\beta\gamma\delta$ & 2013-05-17 12:00 \\
    M3.5 & 2015-02-09 22:19 & 12280 & No & $\beta\gamma\delta$ & 2015-02-08 02:00 \\
    M1.4 & 2014-03-11 16:20 & 12002 & No & $\beta\gamma\delta$ & 2014-03-12 11:00 \\
    M1.1 & 2013-10-14 22:49 & 11861 & No & $\beta\gamma$ & 2013-10-13 09:36 \\
    C8.6 & 2014-08-23 17:19 & 12146 & No & $\beta$ & 2014-08-24 01:00 \\
    C8.6 & 2015-01-06 05:16 & 12253 & No & $\beta\gamma\delta$ & 2015-01-05 01:00 \\
    C8.2 & 2014-12-22 21:15 & 12241 & No & $\beta\gamma\delta$ & 2014-12-21 16:00 \\
    C8.1 & 2015-06-24 15:12 & 12371 & No & $\beta\gamma\delta$ & 2015-06-23 17:00 \\
    C7.9 & 2013-10-11 03:59 & 11861 & Yes & $\beta$ & 2013-10-13 09:36 \\
    C7.6 & 2015-12-13 08:25 & 12468 & No & $\beta$ & 2015-12-13 13:00 \\
    \hline
    \end{tabular}
    \label{tab:events}
\end{table}

\subsection{Coronal Field Extrapolations}\label{sec:extrap}
Extrapolations were conducted at a 12 minute cadence for 6 hours before and 1 hour after a flare event and for 7 hours for flare quiet time windows. NLFFF methods used here to extrapolate the coronal magnetic field follow the same procedures as outlined in \cite{Yurchyshyn2022} and \cite{Garland2024} through the freely available IDL-based \textit{GX Simulator} package (\cite{Nita2015}; \cite{Nita2018}). Magnetograms from the Helioseismic and Magnetic Imager (HMI) are used for the NLFFF extrapolations as the photospheric boundary condition, where the HMI data was rebinned to 1 Mm pixel scale to transform to a local, Cartesian coordinate system  \cite[CEA projection,][]{Thompson2006}. This coordinate transformation allows for the line of sight projection of the HMI base maps onto their line of sight counterparts with minimum distortion by allowing the reference point to be exactly centered on the Cartesian box base. HMI data was not checked for magnetic transients, or brief flare-induced artifacts due to strong Doppler shifts or line reversals in the line-of-sight magnetic field near the flare peak time \cite[e.g.,][]{Sun2017}. However, since the analysis and statistics presented here are investigated hours prior to the flare peak time, any pre-flare effects due to transients are remote. The NLFFF pipeline computes an initial 3D potential field extrapolation, which is used as an initial condition for the NLFFF computation. The NLFFF extrapolation used here follows a weighted optimization method in \cite{Fleishman2017}, which stems from the approach defined in \cite{Wheatland2000}. \textit{GX Simulator} provides the numerical realizations to create the 400 $\times$ 200 $\times$ 100 Mm boxes, which ensure that some of the largest active regions are fully captured. 

\subsection{Magnetic Field Parameters}\label{sec:params}
\renewcommand{\arraystretch}{3}
\begin{table}[ht!]
\centering
\caption{Summary of the magnetic field parameters utilized in this study and their descriptions, calculation formula, and units.}
\begin{tabular}{|c|c|c|} 
 \hline
 \textbf{Description} & \textbf{Formula} & \textbf{Units} \\ 
 \hline
 Total Free Energy & $E_{Free} = \sum \frac{B^2}{2\mu_{0}}dV - \sum \frac{B_{Pot}^{2}}{2\mu_{0}}dV$ & J \\
 Total Current & $J = \sum |\frac{1}{\mu_{0}}\nabla \times \textbf{B}|dA$ & A\\
 Mean Shear Angle & $\Psi = \frac{1}{N} \sum arccos\left( \frac{\textbf{B}\cdot \textbf{B}_{Pot}}{|B||B_{Pot}|} \right)$ & Degrees \\
 Total Twist & $T_{w} = \sum \int_{L} \frac{\mu_{0}J_{||}}{4\pi B} dL$ & Turns \\
 \hline
\end{tabular}
\label{table:4}
\end{table}
The magnetic field parameters investigated here are listed in Table \ref{table:4} and follow the formulas of the Space Weather HMI Active Region Patch (SHARP) parameters \citep{Bobra2015}. The parameters are calculated through summing the individual volume elements within the data volumes. Specifically, the free energy ($E_{Free}$), characterizes the energy available for an eruptive event. The current ($J$) indicates curls in the overall field structure. The angle of the field to the PIL, or magnetic shear ($\Psi$), can be a metric for the stress on the magnetic field; areas of larger shear angles ($> 80^\circ$), where the field is almost parallel to the PIL, can become stressed and trigger an eruption due to departing from its potential field configuration \citep{Leka2007}. The magnetic twist ($T_{w}$), or the number of turns two magnetic field lines are wound about each other, can also be indicative of the magnetic stress and is calculated using the code developed by \cite{Liu2016}. 

\subsection{Magnetic Field Isolation}\label{sec:isolation}
To select portions of the magnetic field that are involved with the flare, two routes were employed to isolate different parts of the extrapolation box: \textit{Active Region Field}, ``AR Field", and the \textit{High Current Region}, ``HCR". Field isolation methods are used to reduce noise from portions of the field not involved with pre-flare field dynamics when calculating the parameters listed in Table \ref{table:4}. The HCR region selection method here resembles the Non-Potential Region (NPR) method discussed in detail in \cite{Garland2024}, but has some slight deviations. It is worth noting that the coordinates for both regions are defined at each time step and, therefore, the regions vary through time. Each of the two methods are described in more detail in the following paragraphs. 

\begin{figure}[ht!]
    \centering
    \includegraphics[width=\linewidth]{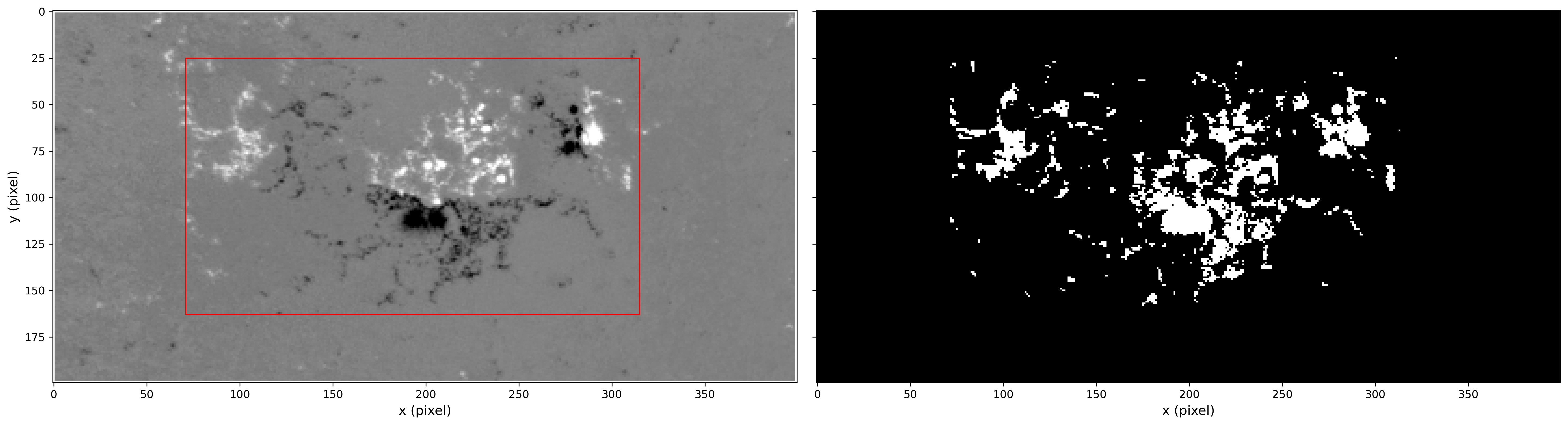}
    \caption{The AR Field isolation bounding box (left) and mask (right) for AR 11748.}
    \label{fig:AR_field_example}
\end{figure}

The isolation of the AR Field is straightforward, which was specifically chosen for its simplicity in any forecasting model. The AR Field is defined as the area where the vertical component of the surface magnetic field, $|B_{z, 0}|$ is $\geq$ 200 G. To create a bounding box around the AR, a smoothing mask was applied to exclude noise from the weak and/or decaying field around the periphery. Then, the bounding box was defined as including $|B_{z, 0}|$ values which were $\geq$ 2 standard deviations above the mean. Coordinates which are within the bounding box and surpass the field strength threshold are used as seed locations to generate the 3D AR Field volume of field lines, from which the parameters are calculated. An example AR Field isolation bounding box and mask (i.e., seed locations) for AR 11748 is shown in Figure \ref{fig:AR_field_example}.

The HCR selection process differs from the AR Field process by utilizing the entire 3D magnetic field volume to isolate sections of interest within the entire active region. To allow automation for the HCR field selection pipeline, regions where the non-potential field, $B_{NP}$, strength distribution were within the largest 0.05$\%$ of the non-potential field distribution were isolated. The non-potential field is calculated as
\begin{equation}
    B_{NP} = B - B_{Pot}.
\end{equation}

\begin{figure}[htb!]
    \centering
    \includegraphics[width=\linewidth]{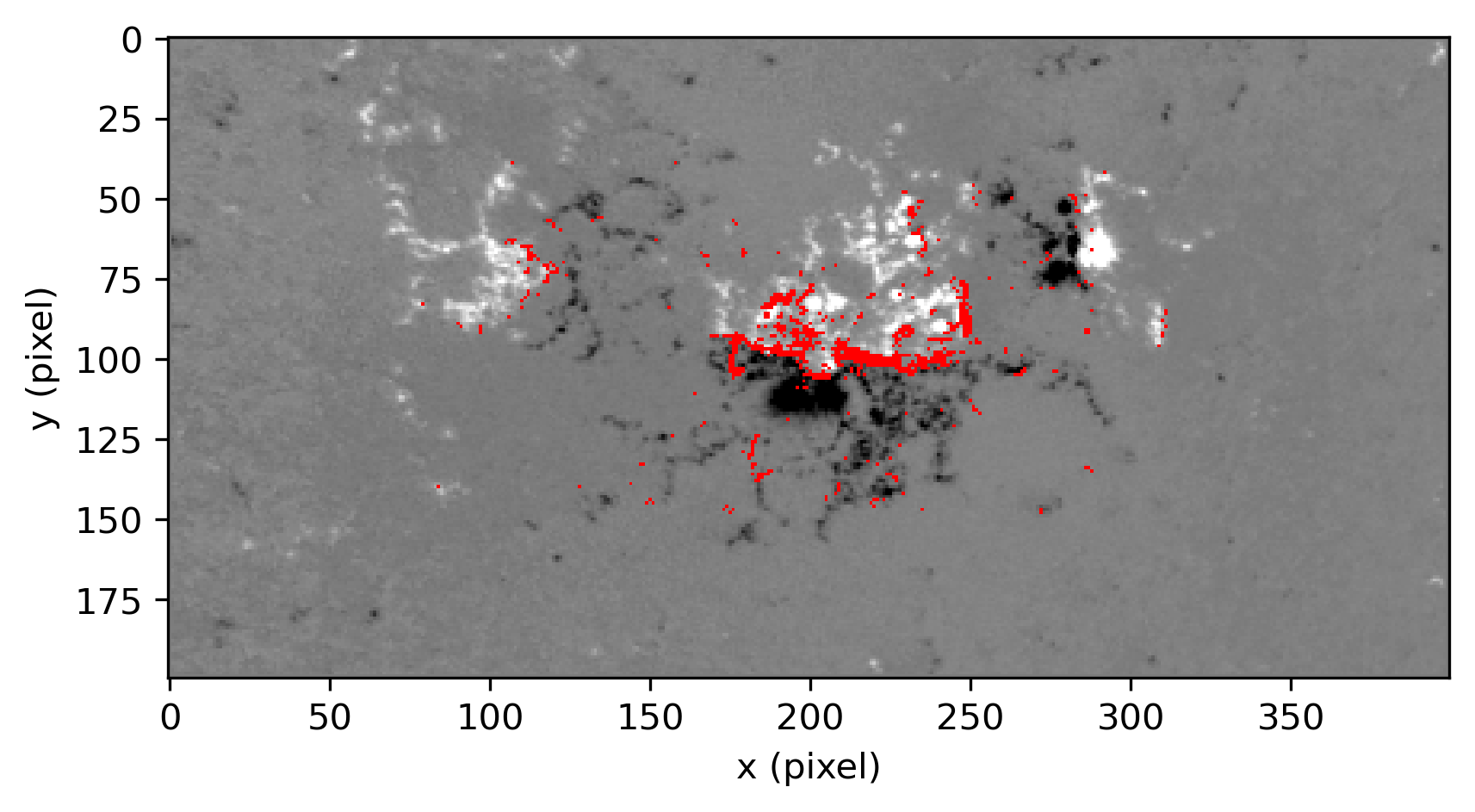}
    \caption{The HCR field isolation for AR 11748.}
    \label{fig:NPR_field}
\end{figure}

\noindent
The HCR then identifies the area within the regions of high non-potential field above the photosphere where $|T_{w}| \geq 1$ turn and $\Psi \geq 80^\circ$. These thresholds indicate locations in the AR Field bounding box that are highly complex and stressed, which could lead to potential flaring activity. The NPR was shown to overlap well with areas of flare brightening as identified from UV imagery \citep{Garland2024}. Where the HCR modifies the Non-Potential Region approach is by including regions where the largest 0.05$\%$ values of the current distribution are located, and these regions are not limited to the areas where $B_{NP}$, $T_{w}$, or $\Psi$ exceed their thresholds. This modification was made to identify areas that may be tied to flaring portions of the magnetic field, but also to includes areas of localized variation based on conclusions in \cite{Kniezewski2024arxiv}. From the 3D coordinates gathered by the HCR method, the field lines generated from the AR Field method, which pass through the HCR in the corona, are traced to their footpoints at the photosphere (see Figure \ref{fig:NPR_field}). These footpoints are used to generate the 3D field line HCR volume, and the parameters are calculated directly from the HCR volume elements.

The analysis discussed in the following sections was performed over all 18 flares listed in Table \ref{tab:events}. All statistical analysis here has particular interest in the time series evolution of the magnetic field parameters in order to gain a deeper understanding into the long term, short scale variability in EUV emission prior to flares presented in \cite{Kniezewski2024arxiv}. To determine if any temporal patterns exist for the individual parameters, a superposed epoch-analysis was performed. Using similar methods to \cite{Kniezewski2024arxiv}, standard deviation ratios between the flaring and flare quiet time periods were calculated at every cumulative hour prior to the flare to assess the pre-flare variability of the magnetic field parameters. Then, the parameter values over height through the corona were analyzed to examine any relationship between the altitude of the magnetic field and its parameters for flaring and flare quiet time windows.

The different forms of analyses address the 18 flares as a whole, and divide the flares by GOES class for more in depth comparisons which are related to flare strength. While significant discussion and conclusions can be taken from the statistics shown below, the reader should keep in mind that the event pool here is small, and there are only 6 flares per GOES class presented here when interpreting these results.

\section{Results}\label{sec:results}

\subsection{Time Series}\label{sec:time_series}
\begin{figure}[!htb]
    \centering
    \includegraphics[width=\linewidth]{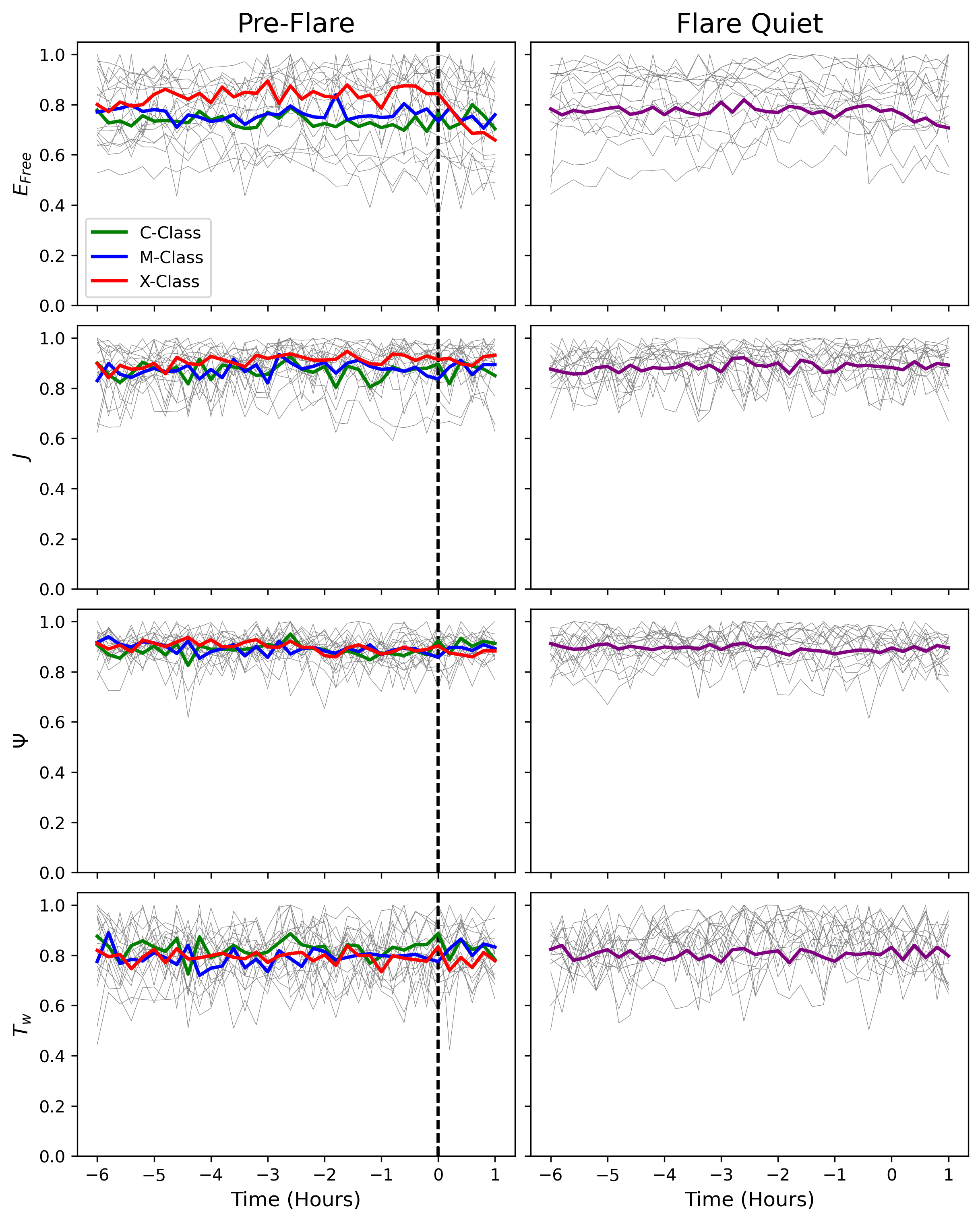}
    \caption{Time series for the four parameters calculated from the AR Field data volume for the pre-flare (left) and flare quiet (right) magnetic fields. Each time series is normalized to the maximum value of the individual series. The black, vertical line indicated the flare start time. The pre-flare time series are averaged at each time step by GOES class, and the flare quiet time series are averaged over all 18 events.}
    \label{fig:AR_timeseries}
\end{figure}

\begin{figure}[!htb]
    \centering
    \includegraphics[width=\linewidth]{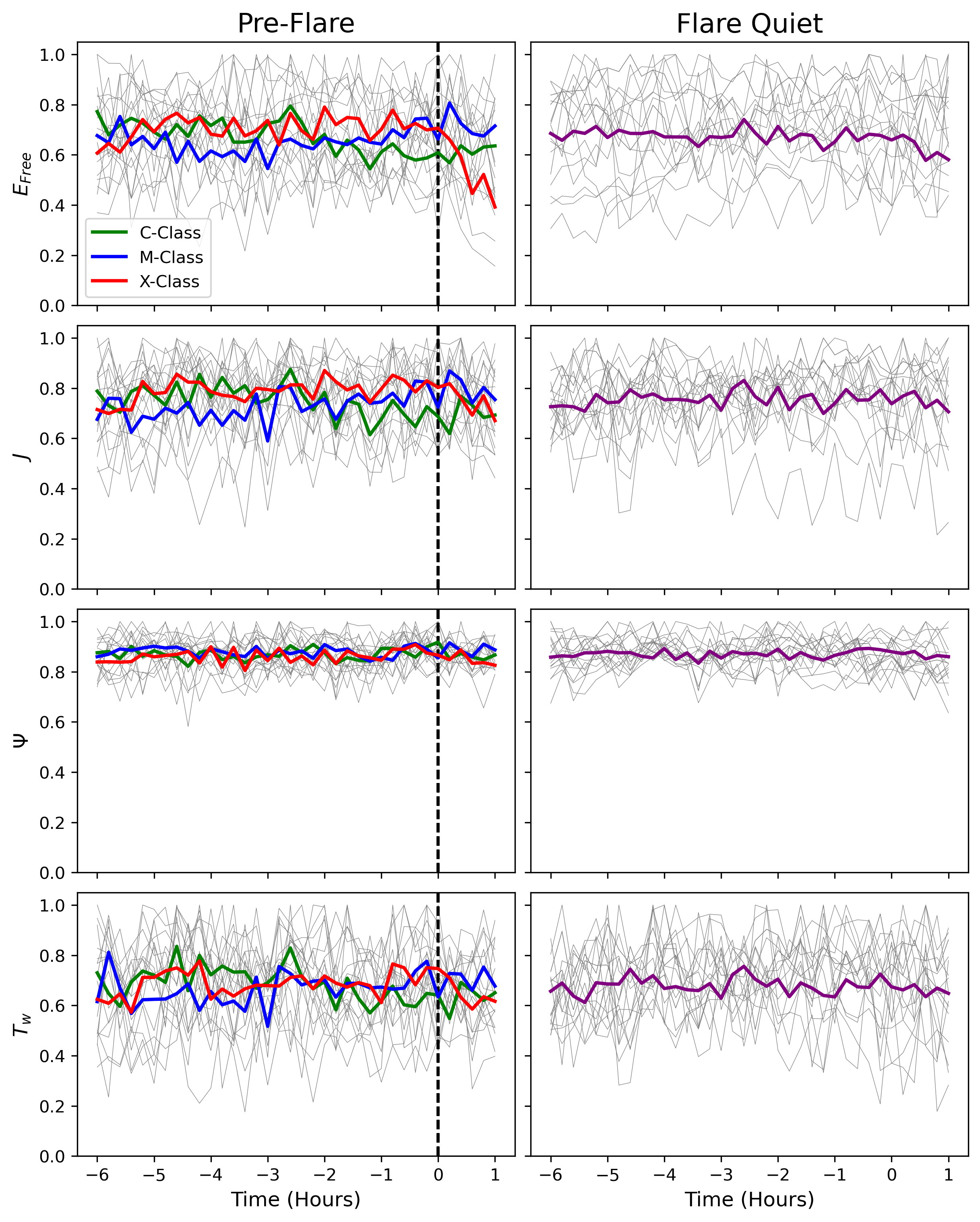}
    \caption{Time series for the four parameters calculated from the HCR Field data volume for the pre-flare (left) and flare quiet (right) magnetic fields. Each time series is normalized to the maximum value of the individual series. The black, vertical line indicated the flare start time. The pre-flare time series are averaged at each time step by GOES class, and the flare quiet time series are averaged over all 18 events.}
    \label{fig:NPR_timeseries}
\end{figure}

Superposed epoch plots are used to analyze any form of consistency among a population time series by averaging the individual time series. Here, superposed epochs were computed for events divided by GOES class to investigate consistencies across events of similar strength. Since the 18 flares here are all centered at the start time of the flare, Figures \ref{fig:AR_timeseries} and \ref{fig:NPR_timeseries} show the superposed epochs for each of the calculated parameters from the AR Field and HCR for the 6 hours prior to and 1 hour after each flare start time for the pre-flare and flare quiet windows. In order to allow for a direct comparison between different events and to be consistent with \cite{Kniezewski2024arxiv}, the parameters were normalized to the maximum of the individual time series so that the dataset ranged only from 0 to 1. 

The most obvious time series trends come from the post-flare relaxation of the magnetic field. The free energy drops dramatically for X-class flares by 22$\%$ in the AR Field and 45$\%$ in the HCR following a flare on average. Decreases in post-flare free energy are most visible in the HCR for individual events, where drops range from 24$\%$ to 64$\%$. Similar drops in post-flare free energy, between $\sim40-60\%$, have previously been reported \cite[e.g.,][]{Aschwanden2019, Garland2024}. Since all the X-class flares in this dataset were associated with a CME (see Table \ref{tab:events}), a significant drop in the free energy is expected, since the free energy available to an active region is the energy source which is released with the flare and an ejecting flux rope. The current and twist calculated from the HCR also show a significant decrease in the hour after stronger flares, with an average decrease around 17$\%$ after X-class flares. However, we emphasize that significant conclusions should not be taken from such small percentage drops of the current or twist. Given the variable nature of the modeled magnetic field, it cannot be said whether this drop has any relation to post-flare dynamics. While any input from instrument noise or error to post-flare trends cannot be neglected, the AR Field and HCR magnetic field isolation methods, which isolate strong portions of the magnetic field by magnitude, were specifically used to remove unwanted noise from the parameter calculations. The upper-bound for noise in the HMI 720s series is reported to be $\sim$6.3 G \citep{Liu2012}, which is well below the thresholds for the AR Field and HCR isolation requirements discussed in Section \ref{sec:isolation}.

Overall, there are no glaring pre-flare trends between any of the parameters across the 18 flare events in the AR Field or HCR field isolation methods. Due to the dynamic and short scale nature of the solar magnetic field, however, trends in the magnetic field have been historically difficult to reveal \cite[e.g.,][]{Leka2003}. Similar conclusions were found in \cite{Kniezewski2024arxiv} for the EUV emission time series, where results ultimately were found consistently in pre-flare variability rather than the temporal trends of the emissions. The magnetic field temporal variability is investigated in the following sections.

\subsection{Standard Deviation Ratios}\label{sec:std_dev_ratio}

\begin{figure}[h!]
    \centering
    \includegraphics[width=\linewidth]{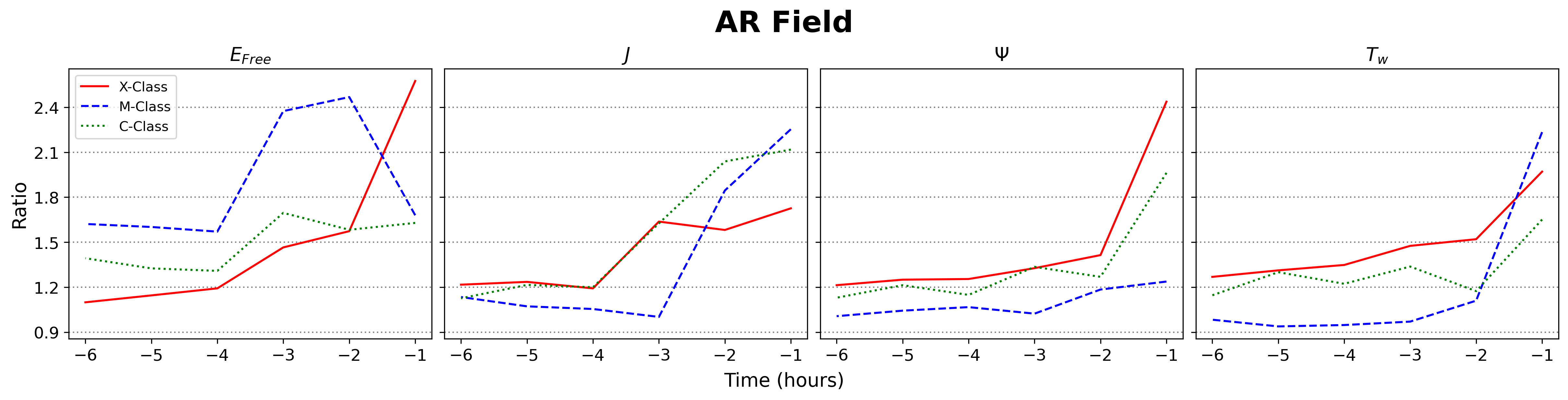}\\
    \includegraphics[width=\linewidth]{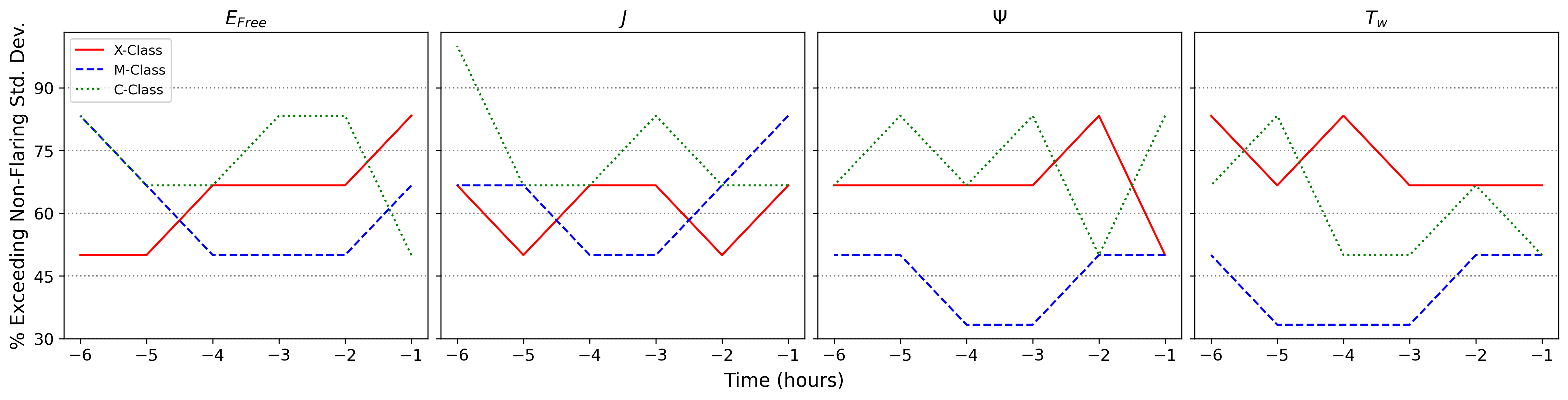}\\
    \includegraphics[width=\linewidth]{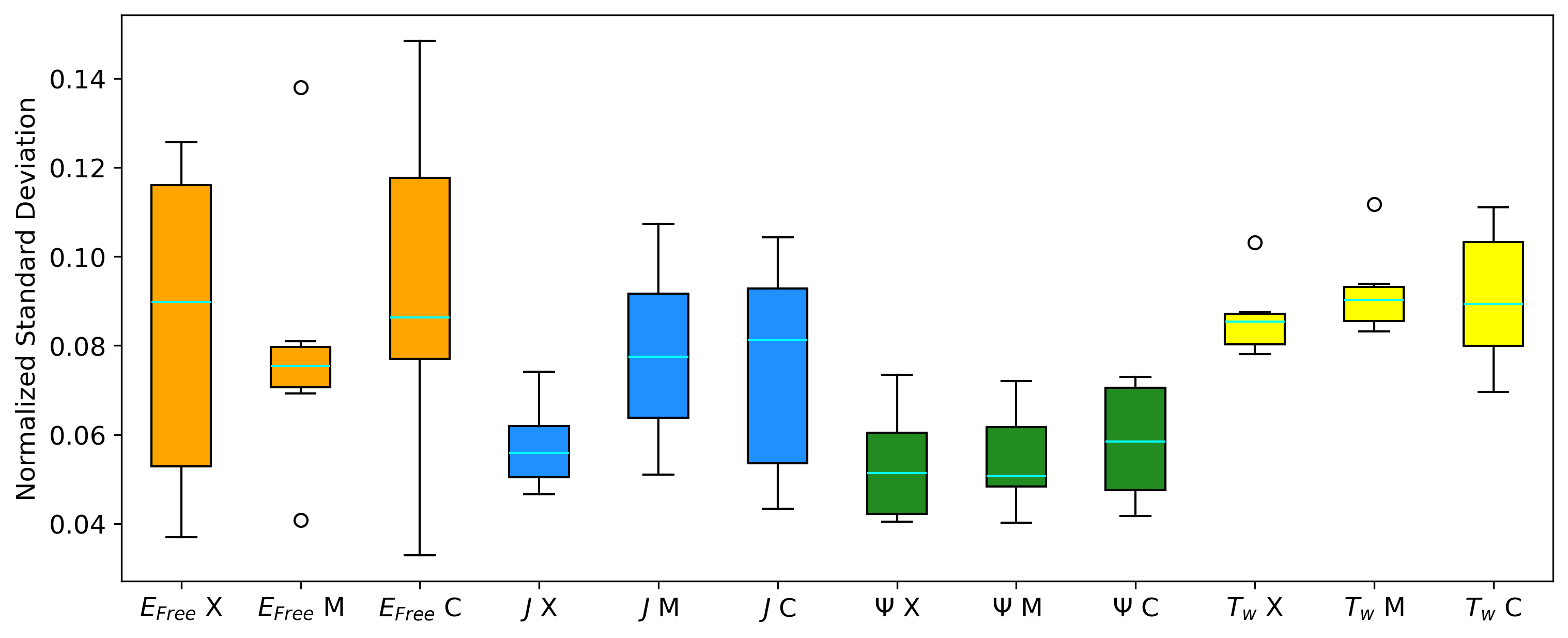}
    \caption{Top: Cumulative hourly standard deviation ratio to flare quiet by GOES flare class for the four calculated parameters from the AR Field at each cumulative hour prior to flare onset. Middle: Percent of flaring standard deviations which exceed the flare quiet standard deviations in the cumulative time windows. Bottom: box and whisker plot of the distribution of the normalized standard deviations for each parameter, separated by GOES Class.}
    \label{fig:AR_std}
\end{figure}

\begin{figure}[h!]
    \centering
    \includegraphics[width=\linewidth]{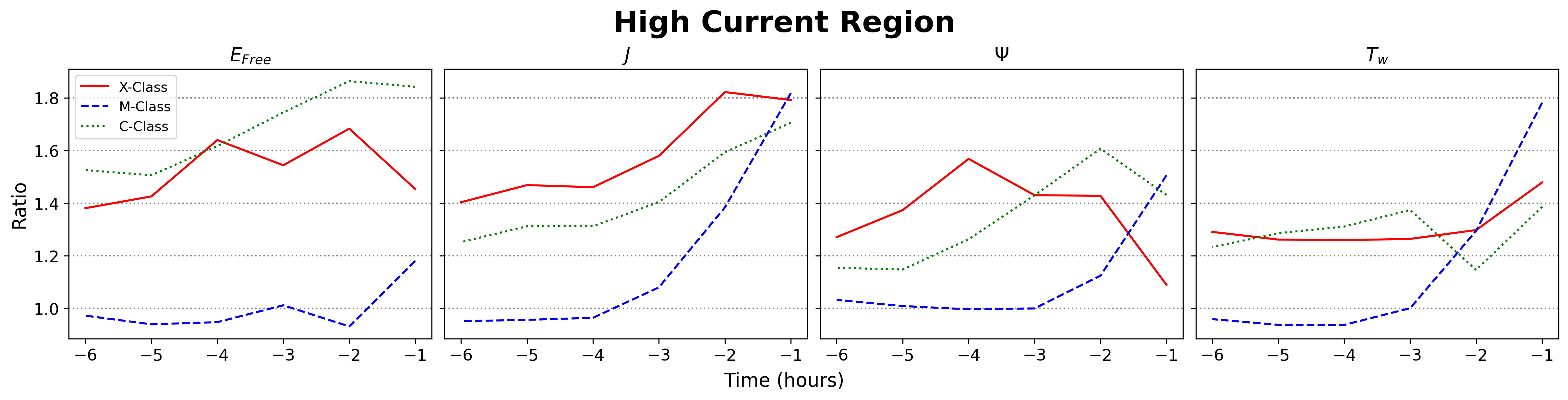}\\
    \includegraphics[width=\linewidth]{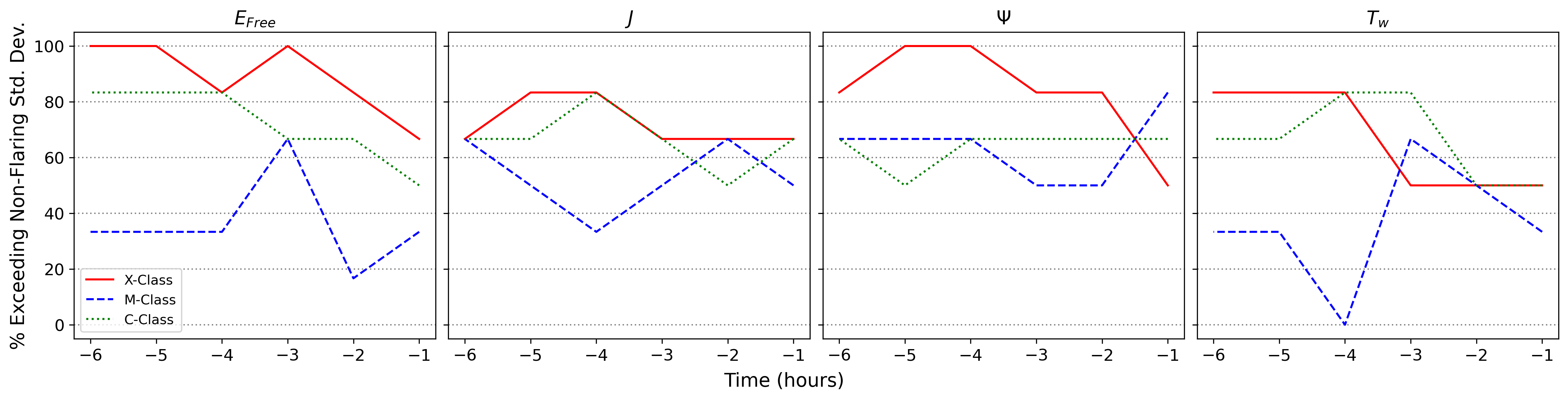}\\
    \includegraphics[width=\linewidth]{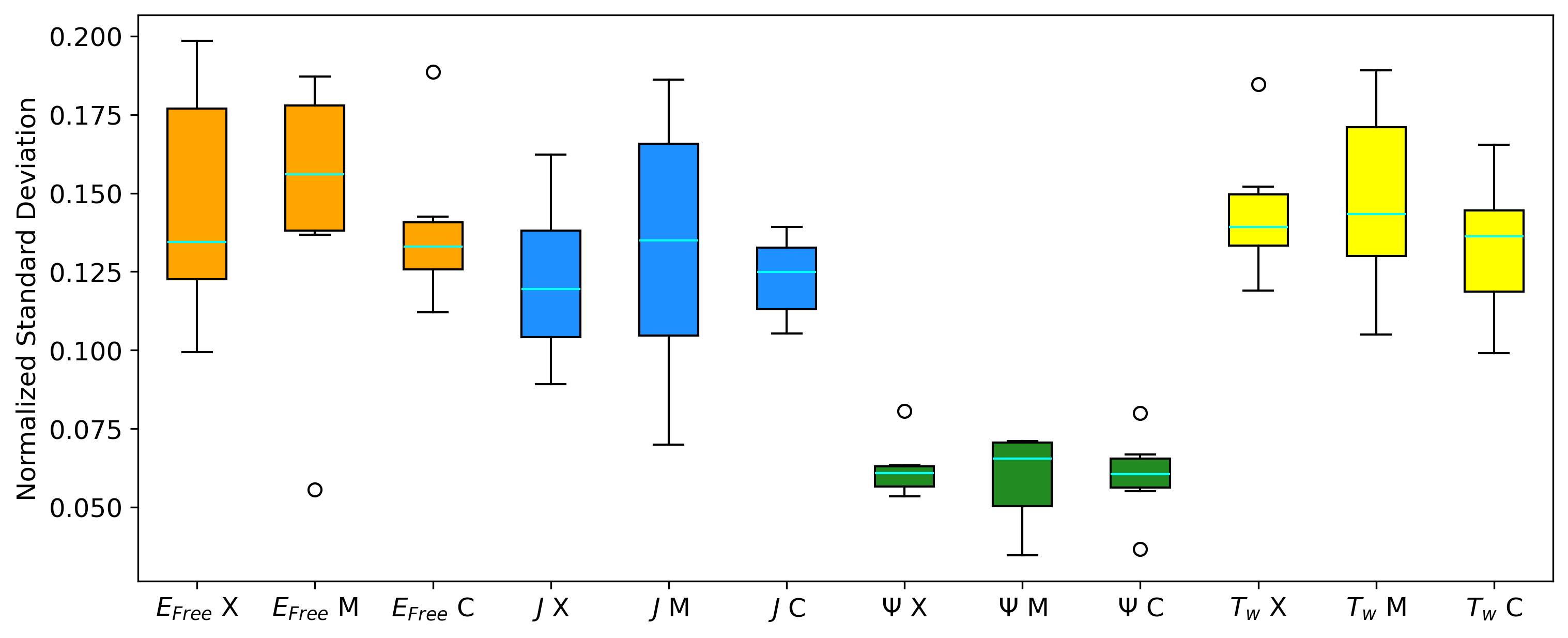}
    \caption{Top: Cumulative hourly standard deviation ratios by GOES flare class for the four calculated parameters from the HCR Field at each cumulative hour prior to flare onset. Middle: Percent of flaring standard deviations which exceed the flare quiet standard deviations in the cumulative time windows. Bottom: box and whisker plot of the distribution of the normalized standard deviations for each parameter, separated by GOES Class.}
    \label{fig:NPR_std}
\end{figure}

A standard deviation ratio was calculated to assess if there is a time period where fluctuations in the parameters are the most apparent prior to the flare. The ratio was calculated similarly to \cite{Kniezewski2024arxiv}, but has some tweaks due to different forms of data availability. Here, the flare quiet windows were taken to be in the \textit{same active region} as the flaring windows. Thus, the flaring time series can be directly compared to the time series when that active region was not flaring. Since observations on the solar limb are limited in observation time due to solar rotation, \cite{Kniezewski2024arxiv} \textit{did not} use the same active regions for the flare quiet comparisons to the pre-flare emissions. A pooled standard deviation, therefore, is not needed here, since the ratio between the pre-flare and flare quiet standard deviations can be taken directly for each flare. The calculation pipeline starts with the population standard deviation, and the ratio is directly calculated using the flare quiet window from the same active region in which the flare occurred. The ratio is calculated for each cumulative hour prior to the flare and is then averaged across each GOES class for each parameter. These results are presented in Figures \ref{fig:AR_std} and \ref{fig:NPR_std} for the parameters calculated from the AR Field and HCR locations, respectively. The box and whisker plot at the bottom of each figure shows the distribution of the normalized standard deviations divided by the magnetic field parameters for each GOES class. Both the AR Field and HCR methods have a small spread of standard deviations, demonstrating that the ratios are not dominated by a singular, highly variable event. Note that the box and whisker y-axis range in \cite{Kniezewski2024arxiv} is set to 0 to 1 to emphasize the small spread standard deviations; Figures \ref{fig:AR_std} and \ref{fig:NPR_std} emphasize this same point, but reduce the y-axis range to display the range of normalized standard deviation distributions for each parameter. Since the standard deviations are similar in variation trends across all cases, the cumulative ratios prior to flare onset is valid and present results consistent across each of the individual flares.

In both the AR Field and HCR methods, the ratio for the current shows a consistent increase in hours leading up to flare onset. The X- and the C-class have a gradual increase for the whole 6-hour time period, and the M-class begin their increase $\sim$4 hours prior to the flare. The free energy ratio is always larger than the flare quiet ratios for the AR Field; the X-class have a rapid increase $\sim$4 hours before and the M-Class peak 2 hours before, with their ratio exceeding 2. In the HCR, the free energy ratio gradually increases prior to onset for the X- and C-class, and the M-class ratio increases $\sim$2 hours prior to flare onset. The shear ratio shows a sharp increase in the 1–2 hours prior to flare onset for the AR Field X- and C-class flares, and a gradual increase for the M-class in the $\sim$3 hours prior to the flare. In the HCR, the X-class shear ratio interestingly decreases for the 4 hours leading up to the flare, but the ratios are all still above 1; the M-class increase through flare onset starting 3 hours before, and the C-class gradually increase for the entire period and peak $\sim$2 hours prior to onset. Finally, the AR Field twist ratio gradually increases for the entire period for the X-class, and the M- and C-class ratios rapidly increase starting 2–3 hours prior to the flare. In the HCR, the M-class twist ratio show the most dramatic increase.

The middle of Figures \ref{fig:AR_std} and \ref{fig:NPR_std} show the percentage of pre-flare magnetic field parameters which had a larger standard deviation than the non-flaring magnetic field parameters. Since there are only 6 events per GOES class, the reader should take caution in interpreting the rapid fluctuations in the percentage values through the cumulative hours. This statistic is useful to gain general information on the parameters which may be most useful as predictors, but a larger dataset would likely present more consistent percentage values throughout the cumulative hours (e.g., the C- and M-class in \cite{Kniezewski2024arxiv} Figure 7). For all 4 parameters, it is evident that the HCR portions of the field exhibit a higher percentage of pre-flare standard deviation which exceed the flare quiet standard deviations. For the X- and C-class, the HCR percentages for all 4 parameters are consistently between 60-100$\%$. Most notably, the X-class free energy and shear percentages fall between 80-100$\%$ for most of the cumulative hour windows in the HCR. The X- and C-class currents and twist also consistently maintain percentages in the HCR between 60-80$\%$ throughout the cumulative hour windows. The M-class have unusual behavior for both the AR Field and the HCR, having the lowest percentages across most of the cumulative hour windows; however, the M-class do exhibit percentages of $\sim$60$\%$ for the current and shear in the HCR. The discrepancy in the M-class is likely due to the low number of events part of each GOES class pool, and a future work including more events could investigate these trends in deeper detail.

\subsection{Cross Wavelet Analysis}\label{sec:xwt}
Wavelet transforms are a useful tool for analyzing physical time series by extracting information on trends and periodicity. The Continuous Wavelet Transform (CWT) is applied to a time series as a band-pass filter and returns regions of high power which correspond to statistically significant oscillations in the time series \citep{Torrence1998}. The Cross Wavelet Transform (XWT) finds the regions of common high power and the phase relationship between two time series \citep{Grinsted2004}. The XWT is particularly useful for analyzing time series which may be connected in some way by physical processes. The XWT analysis was performed with the {\tt PyCWT} Python package \citep{pycwt}, with the first-order detrended parameter time series and default computation recommendations from \cite{Grinsted2004}.

\begin{figure}[h!]
    \centering
    \includegraphics[width=\linewidth]{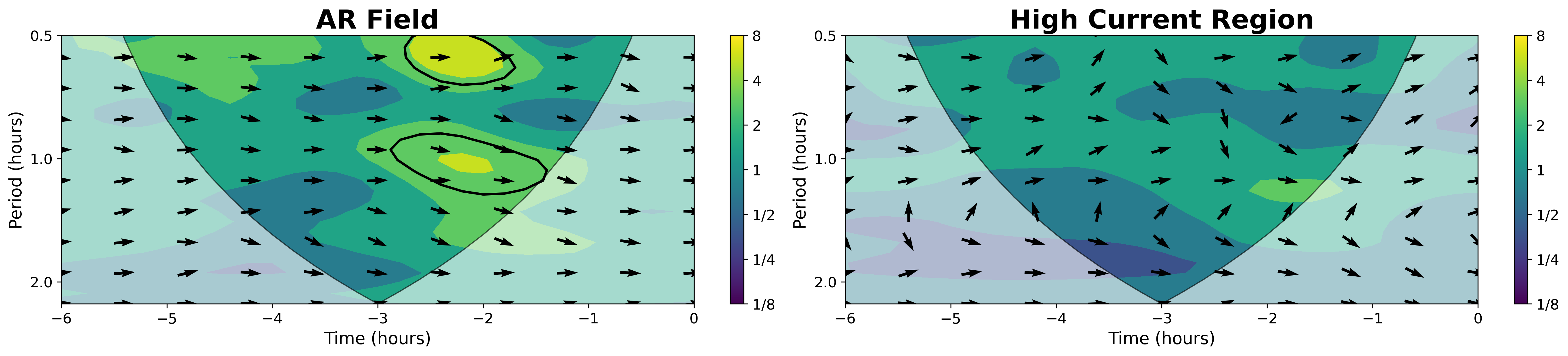}
    \caption{XWT for $J$ and $\Psi$ calculated from the AR Field (left) and HCR (right). The colorbar indicates spectral power.}
    \label{fig:j_vs_shear}
\end{figure}

\begin{figure}[h!]
    \centering
    \includegraphics[width=\linewidth]{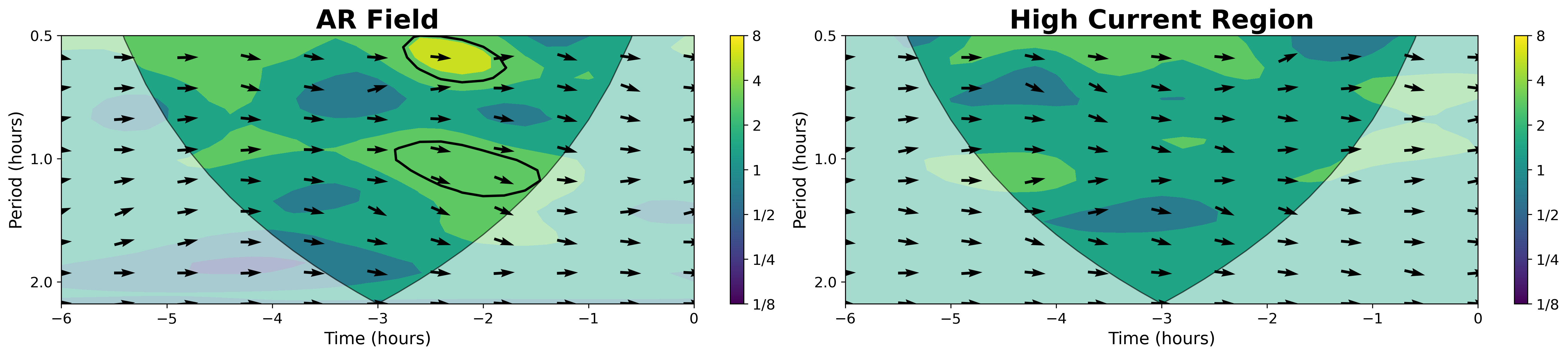}
    \caption{XWT for $J$ and $T_{w}$ calculated from the AR Field (left) and HCR (right). The colorbar indicates spectral power.}
    \label{fig:j_vs_tw}
\end{figure}

\begin{figure}[h!]
    \centering
    \includegraphics[width=\linewidth]{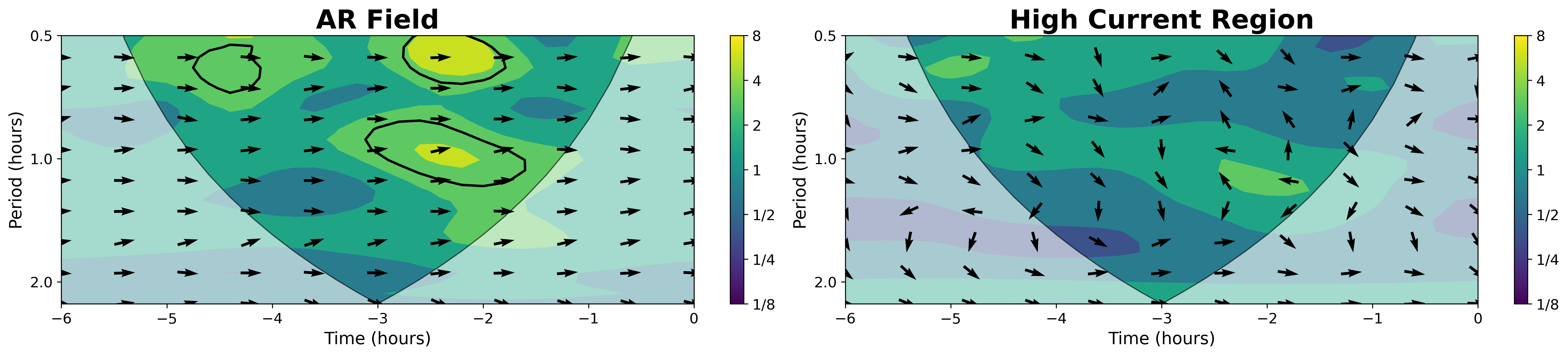}
    \caption{XWT for $\Psi$ and $T_{w}$ calculated from the AR Field (left) and HCR (right). The colorbar indicates spectral power.}
    \label{fig:shear_vs_tw}
\end{figure}

The highest performing XWTs are shown in Figures \ref{fig:j_vs_shear}, \ref{fig:j_vs_tw}, and \ref{fig:shear_vs_tw}, which compare the oscillations between combinations of the current, twist, and shear. In each plot, the left column shows the XWT and phase angle distribution for parameter calculations from the AR Field, and the right shows the same for calculations from the HCR. The XWT y-axis indicates the period of oscillations, and the x-axis is the time in hours before the flare. The color-scaling represents the amplitude of the wavelet power spectrum. The vectors represent the phase angle between the time series, and encircled areas represent oscillations to both time series that are within the 5$\%$ significance level against red noise. The angle of the vectors supplies phase and/or time lag information between the time series; rightward pointing vectors indicate the time series are in phase, leftward pointing vectors indicate the time series are anti-phase, upward or downward pointing vectors indicate there is a time lag between the two time series. Furthermore, an upward or downward arrow both suggest that a time lag exists between them, and the order of each time series in the XWT pipeline determines if the arrow is up or down. Since the wavelet is not localized in time, the black curved lines which begin the gray, shaded region represents the cone of influence (COI), which indicates areas where analysis should be taken with caution due to potential edge effects. 

Since the XWT power spectrum returns a range of values, [$\frac{1}{8}$, $8$] for any analyzed time series, the XWTs in Figures \ref{fig:j_vs_shear}, \ref{fig:j_vs_tw}, and \ref{fig:shear_vs_tw} are averaged XWTs across all 18 flares. An average XWT allows for common powers to two time series to be analyzed for a large pool of events, which may implicate a common time prior to flare onset in which common oscillations occur. Parameters calculated from the AR Field in all 3 XWTs comparing $J$ vs. $\Psi$, $J$ vs. $T_{w}$, and $\Psi$ vs. $T_{w}$ have statistically common oscillations between 1–3 hours before flare onset, with periods between oscillations of 30–60 minutes. The HCR does not show any statistically significant oscillations between any of the time series; this can likely be attributed to the selection process of the HCR. The HCR searches for the areas of highest non-potentiality and highest currents. Since the HCR is redefined at each time step, it may be missing significant oscillations that the AR Field can pick up compared to its background levels. The phase angles for all 3 XWTs and both the AR Field and HCR are all pointing to the right, showing that the time series are all relatively in phase. Since each of these parameters come directly from the magnetic field, their phase relationship is not surprising. Common oscillations between current, twist, and shear prior to a flare do appear to agree with pre-flare theory on magnetic field instability. Variations in twist or shearing would cause spatial variations in the magnetic field, and these spatial changes in the magnetic field may be witnessed through spatial gradients connected to coronal currents. Significant oscillations in the twist and/or shear indicate increasing field instability, which may lead to reconnection and flare onset. 

It is worth briefly discussing the physical mechanism of these oscillations. While we cannot ignore the possibility for instrument effects, it is unlikely the 30-60 minute oscillations are related to HMI artifacts. Currently, only 12 hour and 24 hour oscillations have been reported \cite[e.g.,][]{Smirnova2013}. Additionally, if the oscillations were due to HMI artifacts, the XWT would return a ``band" of high spectral power across the entire observation window, between periods of 30-60 minutes. If these oscillations were due to instrument effects, an isolated area of high spectral power, such as that in Figures \ref{fig:j_vs_shear}, \ref{fig:j_vs_tw}, and \ref{fig:shear_vs_tw}, would not be present.

\subsection{Height Profiles}\label{sec:heightprof}
In addition to addressing the temporal variation of the magnetic field parameters prior to a flare as totals and means over the defined AR Field and HCR volumes, the magnetic field parameters were also inspected through the corona (z-axis) to assess any separation between the parameter quantities as a function of altitude. Historically, the height of flaring portions of the magnetic field, as well as the decay of the field lying overhead, are crucial to many investigations and theory regarding field instability and eruption in the corona \citep{Kliem2006,Torok2004}. 

The plotted height profiles are averaged spatially and temporally. The x- and y-values of the 3D magnetic field parameter volume at each time step were averaged through each 1 Mm increment through the 100 Mm height. Since the standard deviation ratios presented in Section \ref{sec:std_dev_ratio} and XWTs in Section \ref{sec:xwt} for the magnetic field parameters, and the EUV emissions in \cite{Kniezewski2024arxiv} show the most variation at $\sim$2–4 hours before flare onset, the height profiles were averaged across each of the time steps 3 hours prior to the flare. Restricting the time window to only the 3 hours before the flare ensures that the time-averaged height profile is not shifted towards flare quiet conditions. To capture the height profile temporal variation through the 3-hour period, the error bars representing a 95$\%$ confidence interval are included.

To further isolate specific portions of the field, the parameter height profiles were examined for two portions of the field: potential-like field (non-twisted, non-sheared configuration), twisted field ($|T_{w}| \geq 1$), and sheared field ($\Psi \geq 80^\circ$) in both the AR Field and HCR. While the non-twisted, non-sheared field harbors a potential-like configuration, these fields isolated here are \textit{not} current-free. These field lines will be referred to as ``potential" field hereinafter, but the term is used to denote these field lines that have not surpassed the twist/shear thresholds. These field lines likely carry some small degree of twist and shear to allow them to carry currents, but not enough to surpass the twist/shear thresholds. It is emphasized here that in addition to isolating the area defined by the NPR in \cite{Garland2024}, the HCR then modifies this approach by including the field where the current exceeds a threshold. The current region selection is \textit{not} exclusive to only the regions of non-potential field, but can be selected from the entire AR Field box. This brief discussion is to point out that while the NPR does not contain any potential-like field, the HCR does contain non-twisted/non-sheared field lines in addition to non-potential structures. Therefore, isolating the potential-like field from the HCR, and calculating specific parameters from this portion of the field, is possible. 

\subsubsection{Potential Field}\label{sec:potfield}
The height profile which demonstrates the largest separation between pre-flare and flare quiet potential field lines is the vertical current, $J_{z}$. The vertical current is defined as

\begin{equation}\label{eqn_vert_current}
    J_{z} = \frac{1}{\mu_0}\left(\frac{\partial B_{y}}{\partial x} - \frac{\partial B_{x}}{\partial y}\right)dA
\end{equation}

\noindent
from the equation for $J$ listed in Table \ref{table:4}. The vertical component of the current is therefore essentially the curl of the horizontal field. While the 3D magnetic field models does make the total current, $J$, available for analysis, $J_z$ showed greater separation between between pre-flare and flare quiet height profiles. Consequently, discussion of the height profile for $J$ is omitted here. However, it is not surprising that $J_{z}$ appears to be the most dominant among all components of the current; previous studies have found discrimination between flaring and flare quiet active regions from vertical current systems \citep{Wang1994}.

\begin{figure}[h!]
    \centering
    \includegraphics[width=\linewidth]{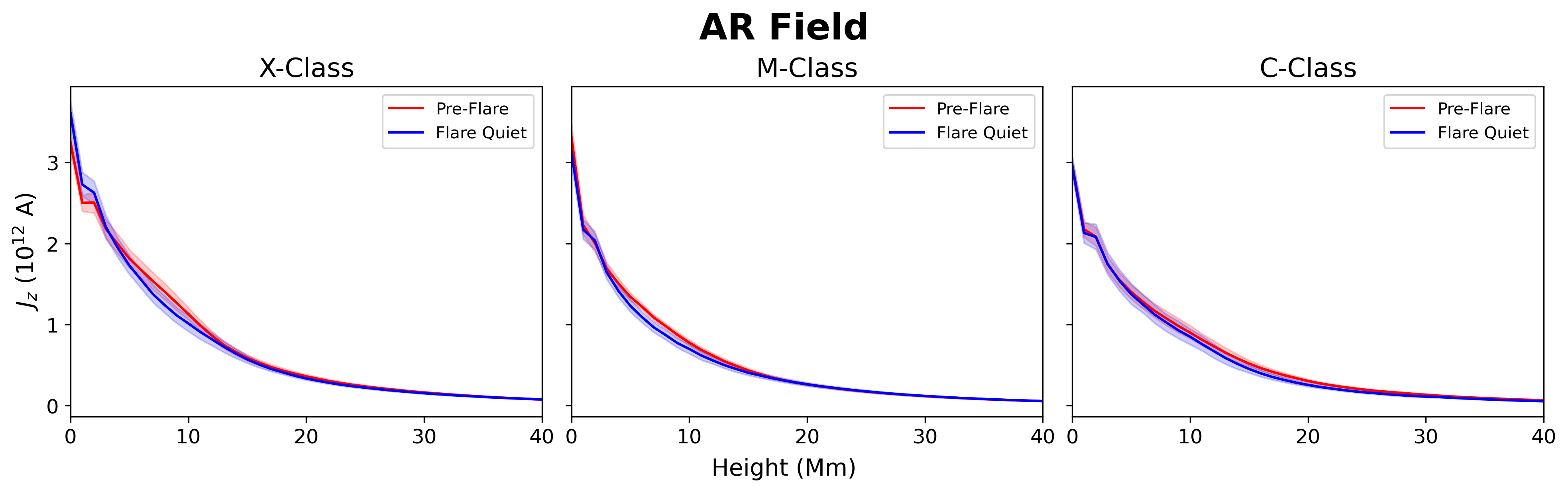}\\
    \includegraphics[width=\linewidth]{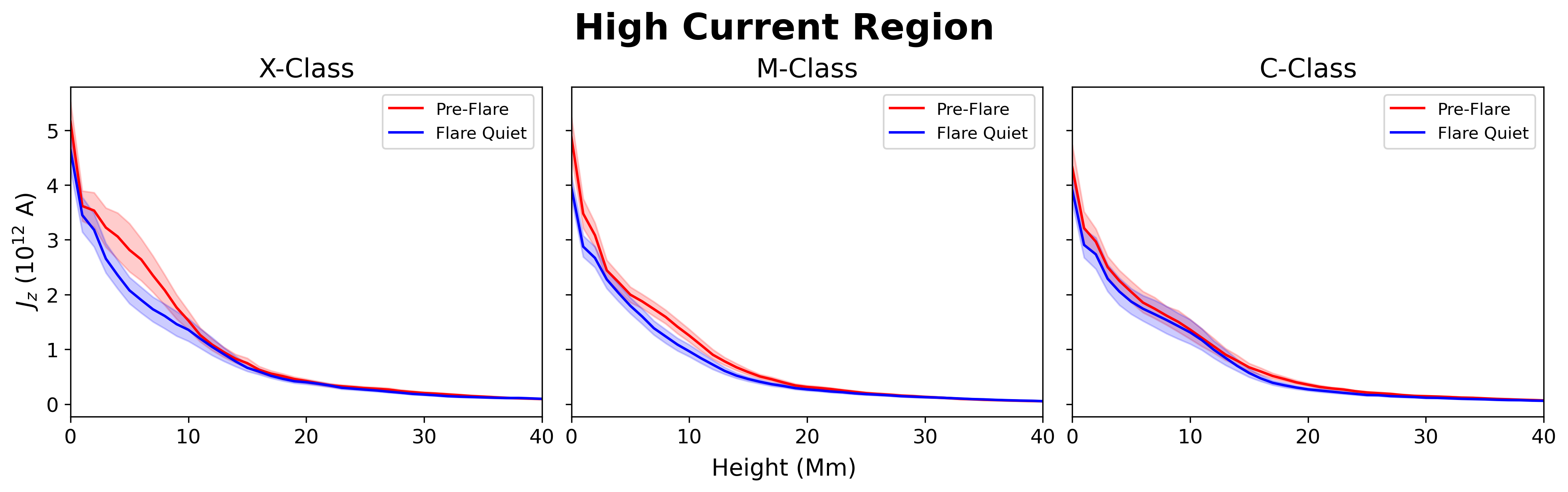}
    \caption{Height profiles for pre-flare (red) and flare quiet (blue) vertical currents, $J_{z}$, isolated to the potential field configurations part of the AR Field (top) and HCR (bottom) regions. The height profiles were averaged by GOES class and the error bars represent a 95$\%$ confidence interval. Note that the x-axis stops at 40 Mm to focus on the lower/mid-corona.}
    \label{fig:pot_vert_current}
\end{figure}

\begin{figure}[h!]
    \centering
    \includegraphics[width=\linewidth]{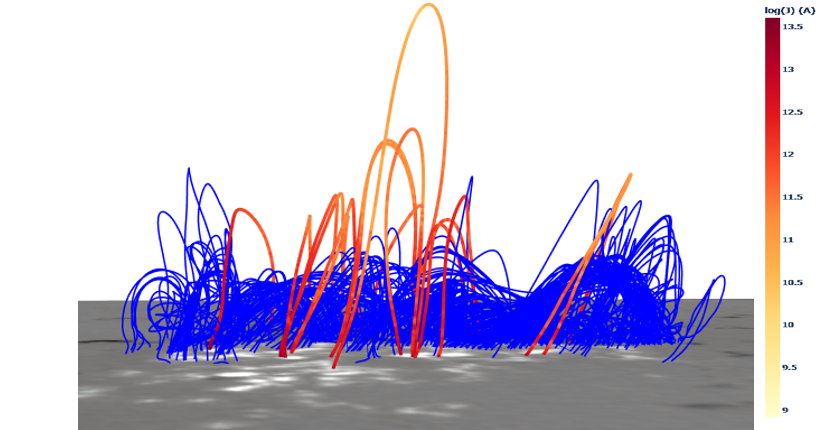}
    \caption{3D Coronal Field volume over the magnetogram for AR 11748. The blue lines indicate lines with $|T_{w}| \geq 1$ or $\Psi \geq 80^\circ$. The red/orange lines represent non-sheared/non-sheared field lines, which are color-scaled to the log of the total current.}
    \label{fig:pot_current_3D}
\end{figure}

The $J_{z}$ height profile in Figure \ref{fig:pot_vert_current} has a significant degree of separation for the HCR. Additionally, it appears that the separation between the pre-flare and flare quiet height profiles seems to scale with GOES class, where the separation becomes larger as flare strength increases. The X-class specifically appears to have a larger confidence interval band than the flare quiet, suggesting that there is greater variation in the pre-flare vertical currents. It is also worth noting that the separation between the pre-flare and flare quiet $J_{z}$ occurs low in the corona ($\sim$0-20 Mm). Figure \ref{fig:pot_current_3D} shows the isolated potential-like field lines which lie over the twisted field lines (blue) for AR 11748, which are color-scaled to the current throughout the field lines. It is evident that the footpoint currents are orders of magnitude higher, and decrease sufficiently with height. 

\begin{figure}[h!]
    \centering
    \includegraphics[width=\linewidth]{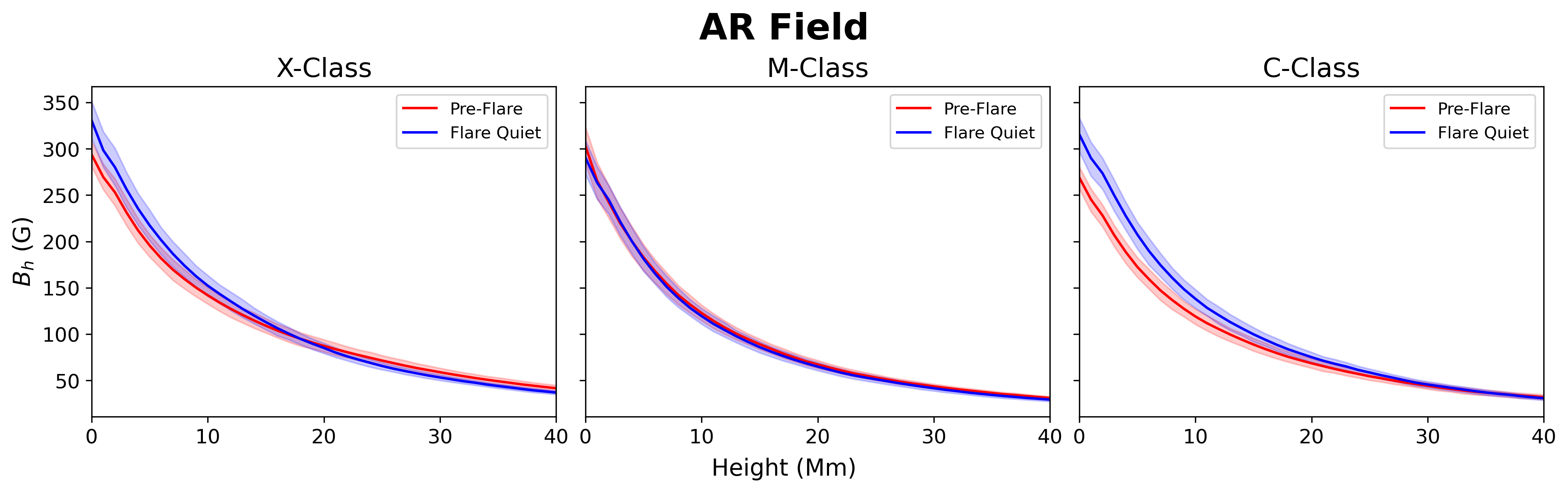}\\
    \includegraphics[width=\linewidth]{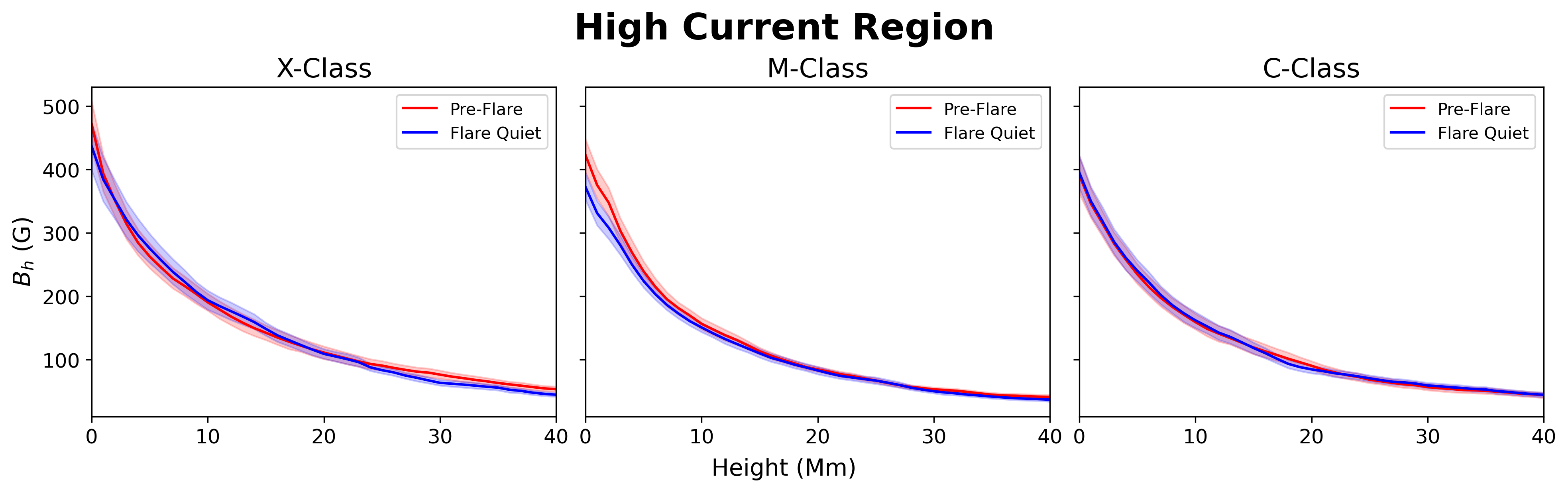}
    \caption{Height profiles for pre-flare (red) and flare quiet (blue) strapping field, $B_{h}$, which are part of the AR Field (top) and HCR (bottom) regions. The height profiles were averaged by GOES class and the error bars represent a 95$\%$ confidence interval. Note that the x-axis stops at 40 Mm to focus on the lower/mid-corona.}
    \label{fig:strapping_field}
\end{figure}

The strapping field is considered to be the potential field that overlies the twisted/sheared magnetic field structures below. The strength of the strapping field through the corona has been extensively studied in relation to CME production following a flare, which can relate to whether an erupting flux rope can escape or not \cite[e.g.,][]{Sun2015}. The strapping field is defined as 

\begin{equation}
    B_{h} = \sqrt{B_{x, Pot}^2 + B_{y, Pot}^2}
\end{equation}

\noindent
where $B_{x, Pot}$ and $B_{y, Pot}$ are the x- and y-potential magnetic field values. The pre-flare and flare quiet strapping field height profiles are shown in Figure \ref{fig:strapping_field}. In the lower corona, there is a slight negative separation for the X- and C-class pre-flare height profiles part of the AR Field, and small positive separation for the M-class flares in the HCR. At higher heights ($\sim$20-40 Mm), there appears to be a minor positive separation between the pre-flare and flare quiet strapping field in both the AR Field and HCR. While a stronger strapping field higher in the corona could indicate, e.g., flux emergence pushing upward on the overlying field, the separation between the pre-flare and flare quiet is only a few Gauss. Future investigation may warrant examining the strength of the strapping field with more events; however, these results show that the difference between pre-flare and flare quiet strapping field strengths through the corona may not be significant. Previous works have also found that the critical height, a parameter used to indicate the decay of the strapping field through the corona \citep{Kliem2004}, does not vary much at all in the hours prior to flare onset \citep{Gupta2024}.

\subsubsection{Non-Potential Field}\label{sec:nonpotfield}
\begin{figure}[h!]
    \centering
    \includegraphics[width=\linewidth]{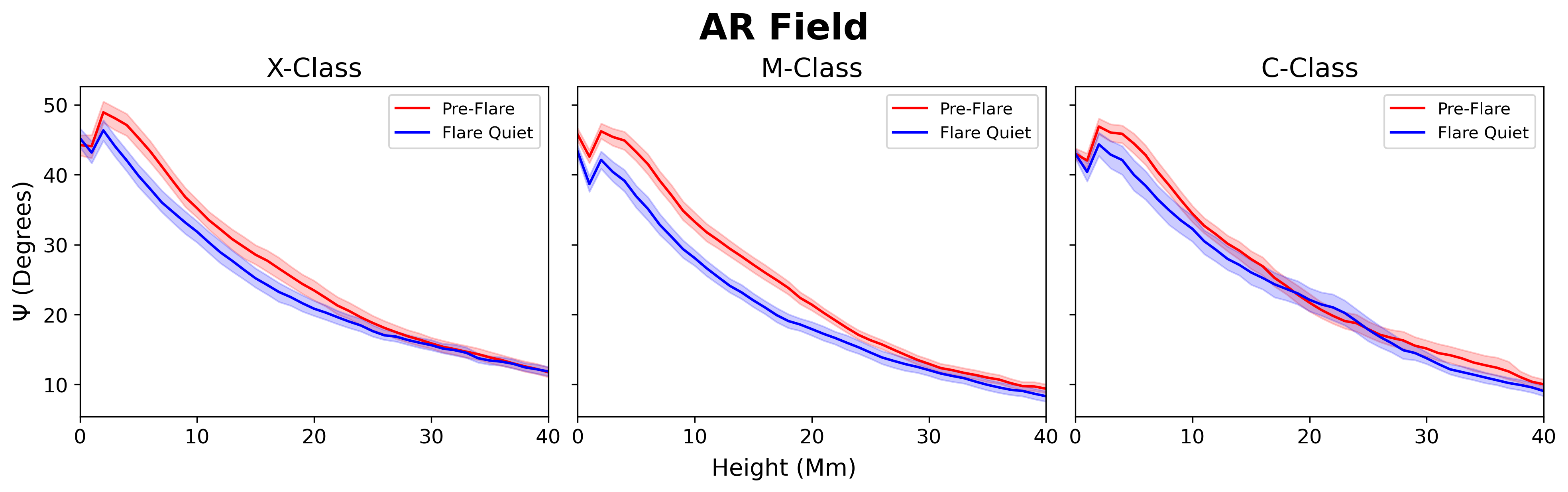}\\
    \includegraphics[width=\linewidth]{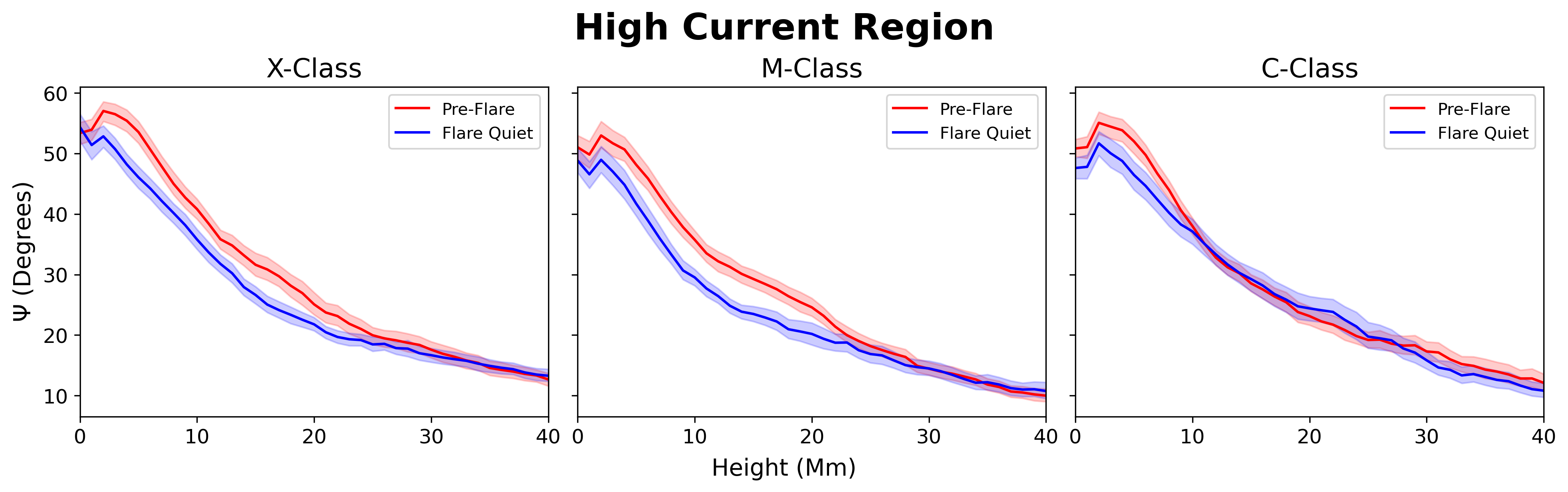}
    \caption{Height profiles for pre-flare (red) and flare quiet (blue) shear angle, $\Psi$, which are part of the AR Field (top) and HCR (bottom) regions. The height profiles were averaged by GOES class and the error bars represent a 95$\%$ confidence interval. Note that the x-axis stops at 40 Mm to focus on the lower/mid-corona.}
    \label{fig:shear}
\end{figure}

The shear height profile was calculated by isolating the field lines in each active region, where $\Psi \geq 80^\circ$ at some section along the field line. From these isolated field lines, the shear angle was averaged over the 100 Mm height through the corona. The height profile of the magnetic shear is shown in Figure \ref{fig:shear}. In both the AR Field and the HCR, there is significant separation between the pre-flare and flare quiet shear height profiles. The separation also persists into higher heights; the average shear angle does not reach flare quiet levels until well above $\sim$30 Mm for the X- and M-class in both the AR Field and the HCR. There also appears to be a dependence on GOES class for both the AR Field and the HCR, where the C-class have much smaller separation and reach flare quiet shear angles at lower heights. 

\begin{figure}[h!]
    \centering
    \includegraphics[width=\linewidth]{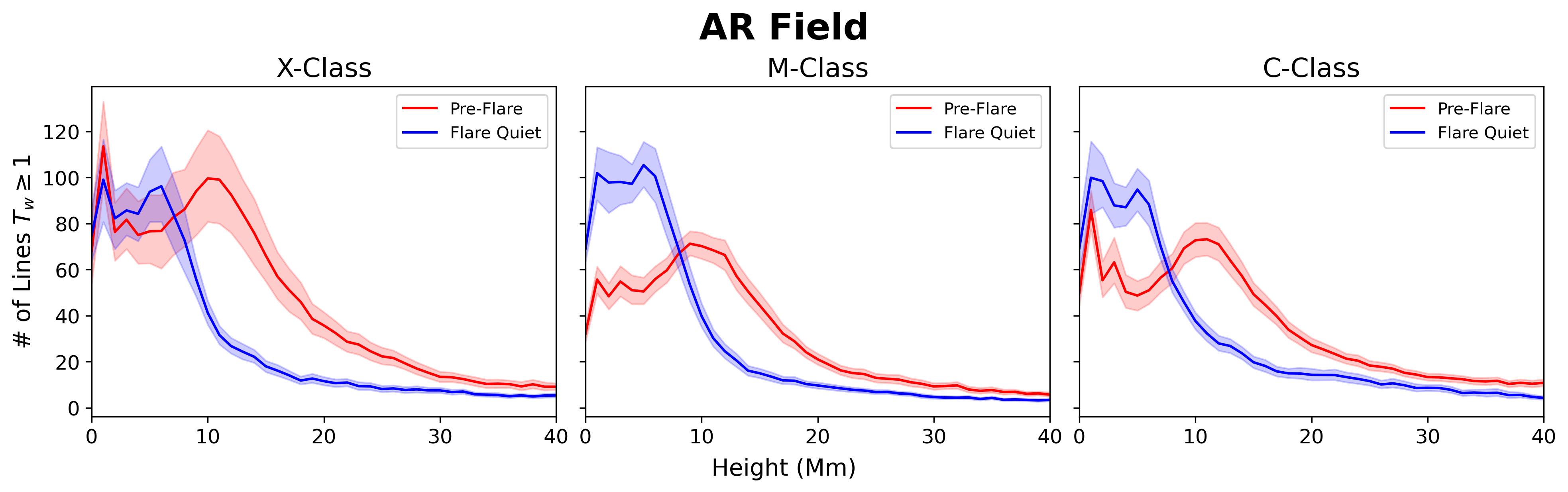}\\
    \includegraphics[width=\linewidth]{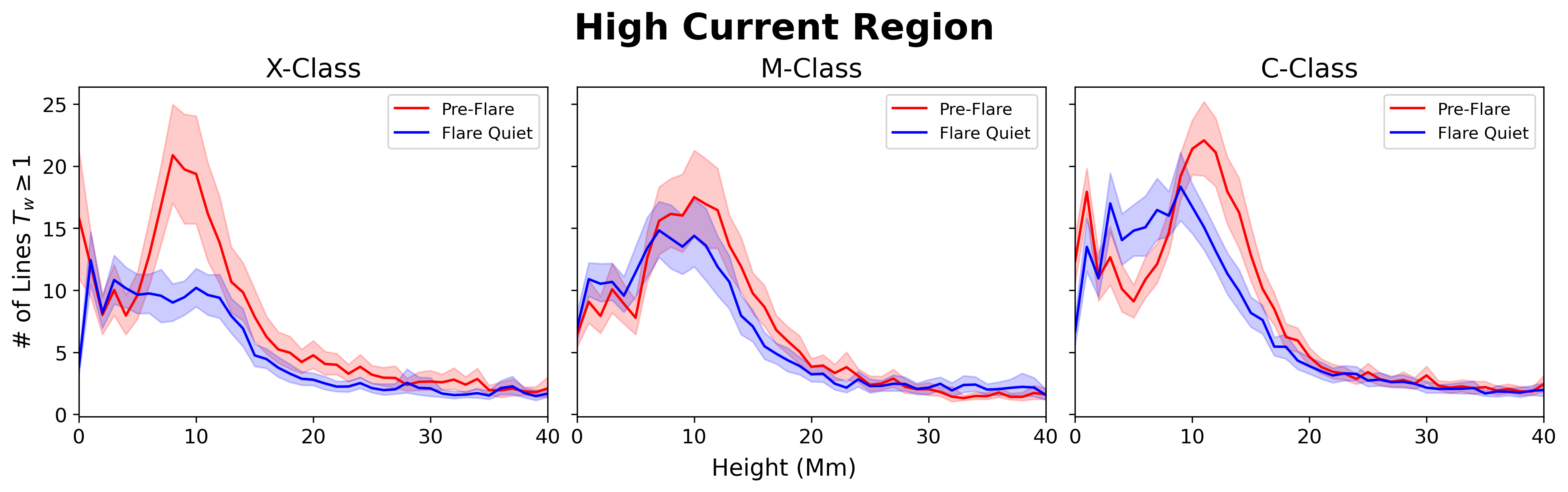}
    \caption{Height profiles for pre-flare (red) and flare quiet (blue) number of field lines which exceed the twist threshold, $|T_{w}| \geq 1$, which are part of the AR Field (top) and HCR (bottom) regions. The height profiles were averaged by GOES class and the error bars represent a 95$\%$ confidence interval. Note that the x-axis stops at 40 Mm to focus on the lower/mid-corona.}
    \label{fig:twist}
\end{figure}

The twist height profile has some small calculation modifications due to how the magnetic twist is computed. Since each field line is assigned a twist number which is given by the turns of the field line, an average twist through the 400 Mm $\times$ 200 Mm x-y plane cannot be computed through the 100 Mm height. Therefore, the twist height profile is modified to isolate the coordinates of the field lines with $|T_{w}| \geq 1$, compute the max height of each field line, and count the number of field lines at each 100 Mm step in which the twist threshold is met. The twist height profile is shown in Figure \ref{fig:twist}. For all GOES classes and both the AR Field and the HCR, there is significant separation between the pre-flare and flare quiet field line count which exceed the twist threshold. As previously discussed, since the pre-flare height profiles are taken from the same active regions during their quiet periods, the separation is not due to vastly different active region magnetic field structures (which might, for example, result from comparing different active regions). While the number of twisted field lines across all GOES classes and in both the AR Field and HCR peak in the lower corona ($\sim$10-20 Mm), all GOES classes do not reach flare quiet numbers until much higher in the corona ($\sim$30-40 Mm). The peak for pre-flare twist is also right-shifted from flare quiet twist, suggesting that the number of twisted field lines increases in height prior to a flare (i.e., building of a magnetic flux rope). A rising flux rope is consistent with the MHD instabilities, where field surpassing twist thresholds for the kink instability and/or passing a height in the corona where the overlying field sufficiently decays with height may have the ingredients for a rapid destabilization. These results indicate that more magnetic field lines are twisted prior to a flare, but that the twisted field lines also persist at higher heights in the corona before flare onset. 

\begin{figure}[h!]
    \centering
    \includegraphics[width=\linewidth]{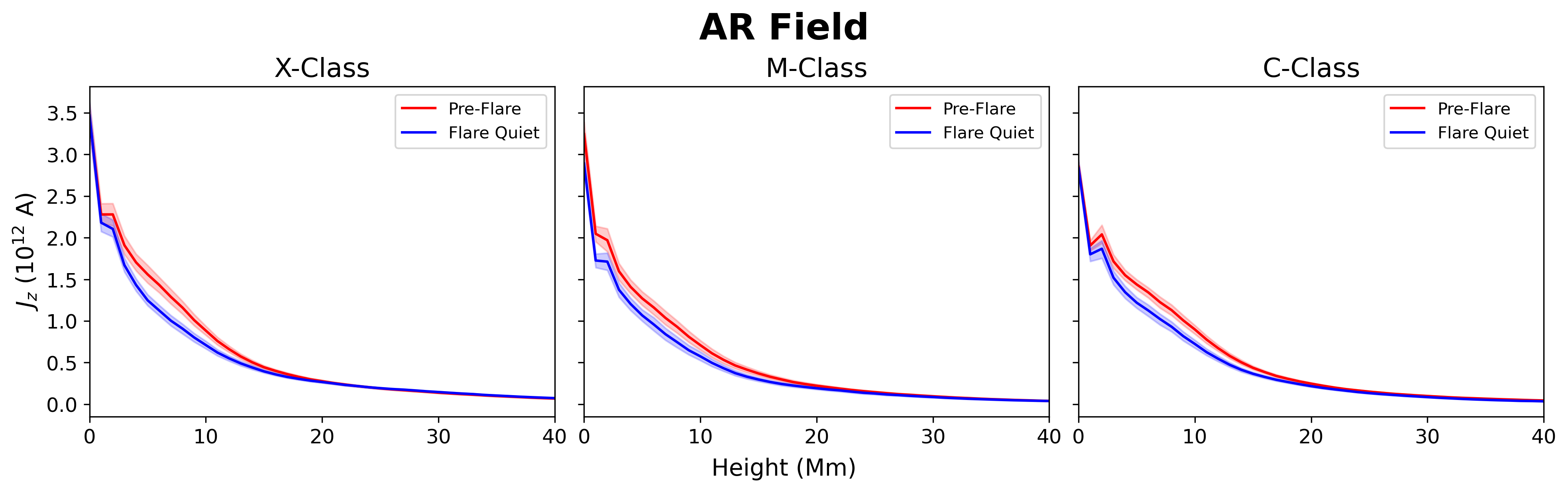}\\
    \includegraphics[width=\linewidth]{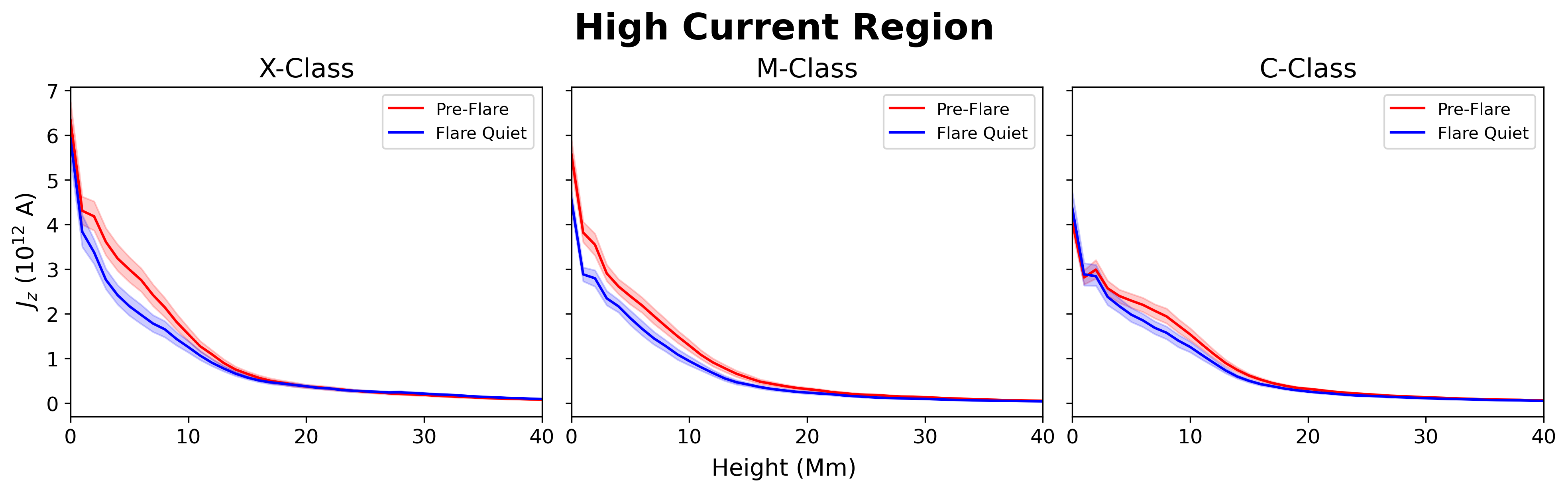}
    \caption{Height profiles for pre-flare (red) and flare quiet (blue) vertical currents, $J_{z}$, isolated to the portions of the field which exceed the twist and/or shear thresholds part of the AR Field (top) and HCR (bottom) regions. The height profiles were averaged by GOES class and the error bars represent a 95$\%$ confidence interval. Note that the x-axis stops at 40 Mm to focus on the lower/mid-corona.}
    \label{fig:tw_vert_current}
\end{figure}

\begin{figure}[h!]
    \centering
    \includegraphics[width=\linewidth]{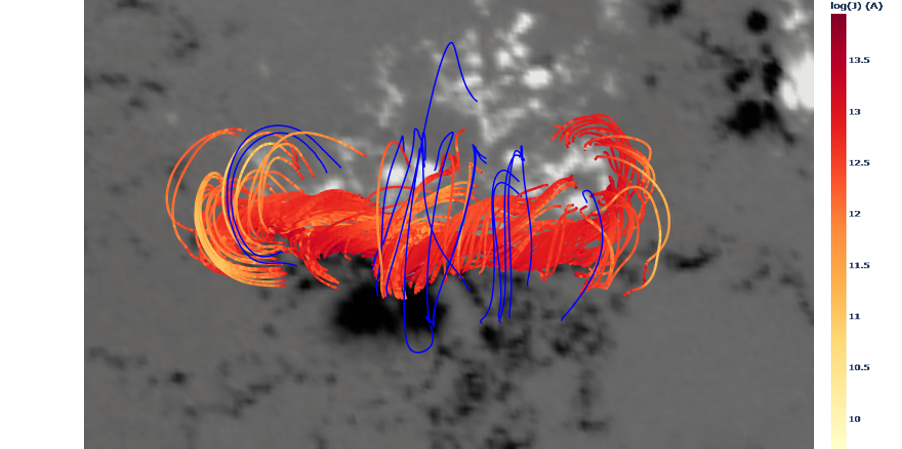}
    \caption{3D Coronal Field volume over the magnetogram for AR 11748. The blue lines indicate lines non-sheared/non-sheared field lines. The red/orange lines represent with $|T_{w}| \geq 1$ or $\Psi \geq 80^\circ$, which are color-scaled to the log of the total current.}
    \label{fig:tw_current_3D}
\end{figure}

To assess the vertical currents part of the non-potential field, the magnetic field lines exceeding $|T_{w}| \geq 1$ and/or $\Psi \geq 80^\circ$ were isolated. The vertical current is calculated using Equation \ref{eqn_vert_current} for the isolated non-potential field lines. The non-potential vertical current height profile is shown in Figure \ref{fig:tw_vert_current}. There is separation between the pre-flare and flare quiet vertical currents in both the AR Field and the HCR, which persists until $\sim$15 Mm into the corona. Additionally, the vertical current separation is noticeably smaller between pre-flare and flare quiet height profiles for the C-class, suggesting that stronger vertical currents may be a component of stronger flares. Figure \ref{fig:tw_current_3D} shows the isolated non-potential field lines, which are color-scaled to the total current, with the potential field lines lying overhead (blue) for AR 11748.

\begin{figure}[h!]
    \centering
    \includegraphics[width=\linewidth]{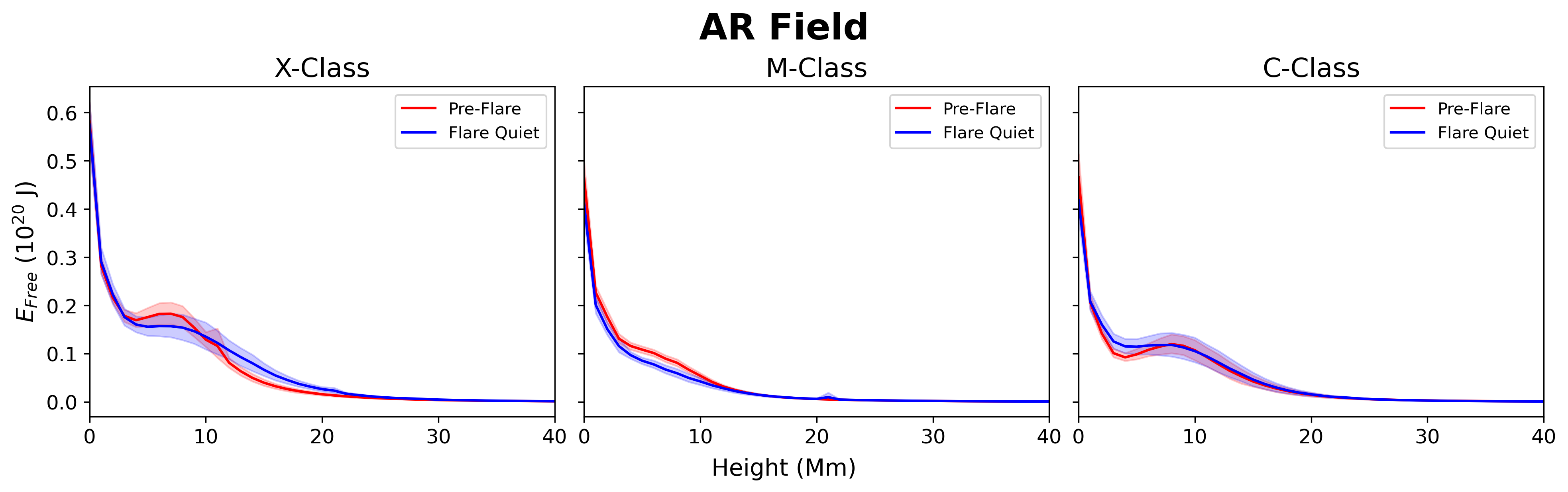}\\
    \includegraphics[width=\linewidth]{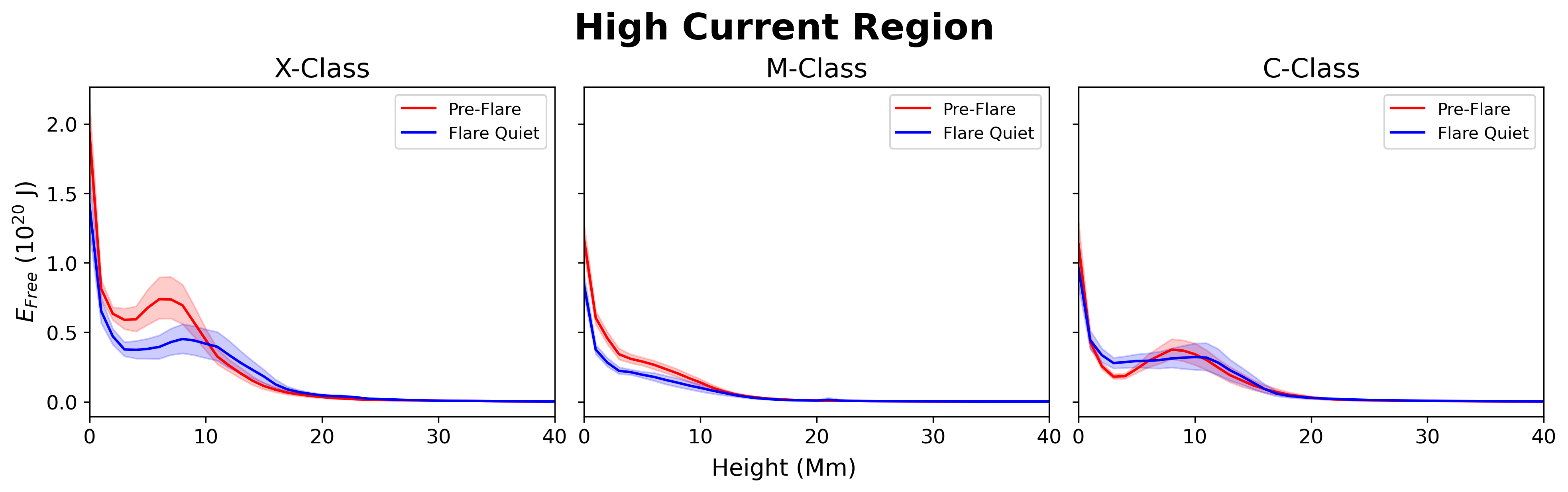}
    \caption{Height profiles for pre-flare (red) and flare quiet (blue) free energy, $E_{Free}$, which are part of the AR Field (top) and HCR (bottom) regions. The height profiles were averaged by GOES class and the error bars represent a 95$\%$ confidence interval. Note that the x-axis stops at 40 Mm to focus on the lower/mid-corona.}
    \label{fig:free_energy}
\end{figure}

Figure \ref{fig:free_energy} shows the free energy height profile. No efforts were made to isolate portions of the potential or non-potential field lines (from, e.g., twist/shear thresholds) from these height profiles; the only forms of isolation used were the AR Field and HCR methods. Since the free energy, by definition, subtracts the potential energy from the total energy (see Table \ref{table:4}), the energy from the larger, potential structures is essentially removed in this quantity. So the free energy height profile is representative of the energy available for flaring or eruption, which consists mostly of non-potential field. In both the AR Field and HCR, there is positive separation for the M- and X-class pre-flare free energy height profiles. It is not surprising that HCR has increasing separation dependent on flare strength, and that the greatest separation exists for the X-class flares. While all the free energy available to an active region is not used in an eruption, the energy available needs to be substantial to power larger events. The significant separation between pre-flare and flare quiet free energy height profiles also provides an interesting discussion on the free energy buildup prior to flare onset. Since the pre-flare height profiles are compared directly to time windows when the same flaring active region is flare quiet, there is a distinct increase in free energy prior to a flare, and this free energy also extends higher into the solar atmosphere. Furthermore, since the flare quiet windows are still part of flare-productive active regions, the free energy is not just a property that remains relatively constant for flaring active regions, but it has variability based on the stress and/or reconfiguration of the magnetic field prior to a flare. 

\subsubsection{Eruptivity}\label{sec:eruptivity}
The relation between the pre-flare height profiles and event eruptivity was also explored. To quantify the difference between the pre-flare and flare quiet height profiles, their 3-hour averaged height profile curves for each parameter and individual event were subtracted from one another. If the mean value of that subtracted curve is positive, then the magnetic field property was stronger through the corona than when the same active region was flare quiet. Since the height profiles appeared to have the most separation in the low corona, the mean was taken for values up to $\sim$20 Mm. Events in the ``high" category indicate that pre-flare height profiles were above its flare quiet height profile curve (i.e., positive separation), and vice versa for the ``low" category. To omit confusion between the potential and non-potential vertical current height profiles, the non-potential vertical current is labeled as ``$NP$ $J_{z}$".

\renewcommand{\arraystretch}{1.2}
\begin{table}[htb!]
\caption{AR Field percentages of eruptive and non-eruptive events of higher/lower height pre-flare vs. flare quiet height profiles.}
\centering
\begin{tabular}{l cc cc}
\hline
 & \multicolumn{2}{c}{High $\%$} & \multicolumn{2}{c}{Low $\%$} \\
\cline{2-3} \cline{4-5}
Parameter & CME & No CME & CME & No CME \\
\hline
$J_{z}$ & 2 (22$\%$) & 7 (78$\%$) & 8 (89$\%$) & 1 (11$\%$) \\
$B_{h}$ & 3 (50$\%$) & 3 (50$\%$) & 7 (58$\%$) & 5 (52$\%$) \\
$\Psi$ & 9 (60$\%$) & 6 (40$\%$) & 1 (33$\%$) & 2 (67$\%$) \\
$T_{w}$ & 7 (54$\%$) & 6 (46$\%$) & 3 (60$\%$) & 2 (40$\%$) \\
NP $J_{z}$ & 8 (57$\%$) & 6 (43$\%$) & 2 (50$\%$) & 2 (50$\%$) \\
$E_{Free}$ & 6 (60$\%$) & 4 (40$\%$) & 4 (50$\%$) & 4 (50$\%$) \\
\hline
\label{table:AR_eruption}
\end{tabular}
\end{table}

\renewcommand{\arraystretch}{1.2}
\begin{table}[htb!]
\caption{HCR percentages of eruptive and non-eruptive events of higher/lower height pre-flare vs. flare quiet height profiles.}
\centering
\begin{tabular}{l cc cc}
\hline
 & \multicolumn{2}{c}{High $\%$} & \multicolumn{2}{c}{Low $\%$} \\
\cline{2-3} \cline{4-5}
Parameter & CME & No CME & CME & No CME \\
\hline
$J_{z}$ & 5 (50$\%$) & 5 (50$\%$) & 5 (62$\%$) & 3 (38$\%$) \\
$B_{h}$ & 4 (36$\%$) & 7 (64$\%$) & 6 (86$\%$) & 1 (14$\%$) \\
$\Psi$ & 9 (64$\%$) & 5 (36$\%$) & 1 (25$\%$) & 3 (75$\%$) \\
$T_{w}$ & 7 (64$\%$) & 4 (36$\%$) & 3 (43$\%$) & 4 (57$\%$) \\
NP $J_{z}$ & 8 (62$\%$) & 5 (38$\%$) & 2 (40$\%$) & 3 (60$\%$) \\
$E_{Free}$ & 8 (62$\%$) & 5 (38$\%$) & 2 (40$\%$) & 3 (60$\%$) \\
\hline
\label{table:HCR_eruption}
\end{tabular}
\end{table}

The percentage breakdown relating to whether a flare was associated with a CME or not and pre-flare height profile curve related to its flare quiet height profile curve is shown in Tables \ref{table:AR_eruption} and \ref{table:HCR_eruption}. For the parameters measured from the non-potential field ($\Psi$, $T_w$, $NP$ $J_z$, and $E_{Free}$) and calculated in the HCR, an increase in their magnitude is almost twice as likely to lead to a CME. A CME is also almost twice as likely to occur when the vertical current in the overlying field is lower than the flare quiet conditions. The largest difference in percentages comes from the strength of the overlying field; active regions that have a weaker overlying field prior to a flare compared to its flare quiet conditions were over six times likely to host a CME. The non-potential parameters part of the AR Field do show a marginal increase in CME production, but the best-performing parameters appear to be connected to the potential field. A flare is over 3 times likely to be confined if the vertical current in the overlying field has a higher height profile than the active region's flare quiet conditions. Additionally, in the AR Field, a flare is 8 times likely to host a CME if the $J_z$ height profile for the overlying field is lower than flare quiet conditions. While a larger dataset is needed to corroborate these results, the interplay between the non-potential and potential field structures and CME production provides an intriguing discussion. CME productivity of a flaring active region is typically associated with the decay of the overlying potential field strength with height. $J_z$ is most prevalent in the twisted/sheared field structures which make up magnetic flux ropes, but there are currents concentrated at lower heights in the corona in the larger non-twisted/non-sheared magnetic field structures (as discussed in Section \ref{sec:heightprof}). Since $J_z$ is calculated by curls of the horizontal field, the interplay between currents and CMEs may be tied to the strength of the curls of the overlying field.

\section{Conclusions}\label{conclusions}
Here, the coronal magnetic field was extrapolated using a NLFFF model for 18 on-disk flares for 6 hours before and 1 hour after the flare start time to investigate pre-flare magnetic field variability. Flares were specifically chosen from active regions which did not have flares $>$ C5.0 within the extrapolation window, ensuring the chosen events are free from saturation due to other flares. Flare quiet periods for each flaring active region were also extrapolated to compare pre-flare quantities to quiet times. The magnetic field was isolated into two different regions to select portions of the field which may be involved with the flare and/or responsible for thermal variation:
\begin{enumerate}
    \item \textit{Active Region (AR) Field} - areas where the surface (photosphere) magnetic field, $|B_{z,0}|$, is $\geq$ 200 G
    \item \textit{High Current Region (HCR)} - areas throughout the 3D field volume which were part of the largest $0.05\%$ current, $J$, and non-potential field strength, $B_{NP}$, of their distributions and surpassed the twist and shear thresholds, $|T_{w}| \geq 1$ or $\Psi \geq 80^\circ$. 
\end{enumerate}
\noindent
The HCR is calculated similarly to NPR isolation method from \cite{Garland2024}, but adds in the regions of high current which are not required to be part of the non-potential field (i.e., exceed twist or shear thresholds). From this method, larger non-twisted/non-sheared fields were included because they were ultimately found to include higher currents. 

The magnetic field parameters calculated here are the free energy, $E_{Free}$, twist, $T_w$, shear, $\Psi$, and current, $J$. Analysis of the time series was the primary focus here to search for any consistencies between the pre-flare on-disk magnetic field parameters variation. Superposed epoch analysis of each parameter were performed to search for any pre-flare trends across all the events. Standard deviation ratios between the pre-flare and flare quiet parameters were calculated at each cumulative hour prior to flare onset to quantify time periods before the flare when the pre-flare parameters were most variable. To correlate which parameters varied at similar time periods prior to flare onset, XWTs compared the time series of parameters prior to a flare and searched for significant oscillations between the two time series. Height profiles of parameters isolated to the potential and non-potential components of the field were utilized to compare the difference between pre-flare and flare quiet parameter strength and height in the corona prior to flare onset.  The main implications are summarized as follows:
\begin{enumerate}
    \item \textbf{Parameters showed consistent decreasing trends following a flare, but no pre-flare trends were found.} Since the 4 parameters calculated here are typically strong in areas of high twist/shear, it is not surprising that there is dramatic decrease in $E_{Free}$ and $T_w$ for stronger flares, which release a significant portion of energy and flux if associated with a CME following a flare. Additionally, a lack of pre-flare trends is somewhat expected. The magnetic fields here and pre-flare EUV emission time series in \cite{Kniezewski2024arxiv} rarely showed any trends or uniformity, which is what led both studies to perform analysis on characterizing the variability of the time series in the hours before flares.
    \item \textbf{All 4 pre-flare parameters vary significantly starting 2–4 hours prior to a flare and increase in variability until flare onset.} This time period agrees well with the emission variation in \cite{Kniezewski2024arxiv} and suggests these methods, in addition to utilization of EUV emission, can provide a window to test for flare forecasting methods. An interesting find here, however, is that the magnetic field parameters start varying 2–4 hours before and increase their variance until the flare start time. The EUV emissions in \cite{Kniezewski2024arxiv} peaked prior to flare onset, and these peaks seemed to scale with GOES class. The reasoning behind this discrepancy is currently unknown. It is likely that developing routines to compare the EUV emission and magnetic fields directly for each event individually could provide more information on the coupling of pre-flare magnetic and thermal dynamics in the corona. 
    \item \textbf{The standard deviation ratios for most of the parameters and GOES classes never falls below 1.} While the XWTs indicate that the most significant oscillations between the currents, shear, and twist occur between 2–4 hours prior to flare onset, \cite{Kniezewski2024arxiv} also reported that the standard deviation ratios in pre-flare EUV emission never fell below 1. The consistency between these two studies may suggest variation in the magnetic field and emission could start several hours prior to flare onset.
    \item \textbf{There is significant separation between the pre-flare and flare quiet parameters through their height in the corona, which has connection to whether or not a flare will be eruptive.} The non-potential parameter ($\Psi$, $T_w$, $NP$ $J_z$, and $E_{Free}$) are consistently stronger through $\sim$20 Mm into the corona, and the separation scales with GOES class. The pre-flare potential field strength is slightly lower, but the potential field vertical currents are significantly stronger close to the footpoints ($<$ 10 Mm). 
\end{enumerate}

The findings presented here support that the multiple components of the magnetic field are exceptionally variable prior to flare onset. Further research expanding this work, with, e.g., a larger pool of events and near-real-time data, is planned to investigate the predictability of these methods. Isolating areas of an active region involved with flaring through the HCR is promising, and future work includes comparing the HCR seed locations with eventual flare ribbons. The HCR is an extension of the NPR introduced in \citep{Garland2024}, who found that the NPR overlapped well with flare ribbons when there was a well-defined polarity inversion line. Forecasting accuracy may improve if the HCR shows consistent overlap with eventual flare ribbons across a variety of active reigon complexity. We emphasize that the methods here and in \cite{Kniezewski2024arxiv} are computationally inexpensive and simple; the coronal magnetic field and EUV emission integrations can both be calculated on the order of minutes for each observation time step. In both studies, the standard deviation ratios show the most consistent trends prior to flare onset; the magnetic fields and EUV emission showed considerable variation 2-4 hours before a flare with $\sim$60-80$\%$ accuracy, and with higher accuracy for stronger flares. By incorporating on-disk observations of the EUV emission in conjunction with the magnetic fields, both data products together can easily be incorporated into an automated flare prediction model in a near-real-time capacity.

\begin{acknowledgments}
The views expressed are those of the authors and do not reflect the official guidance or position of the United States Government, the Department of Defense (DoD) or of the United States Air Force.
\end{acknowledgments}

\bibliography{bibliography}{}
\bibliographystyle{aasjournal}



\end{document}